\documentclass[aps,preprint,nofootinbib]{revtex4-1}
\usepackage{amsfonts}
\usepackage{amsmath}
\usepackage{amssymb}
\usepackage{eurosym}
\usepackage{multirow}
\usepackage{paralist}
\usepackage{graphicx}%
\usepackage[colorlinks=true,linkcolor=blue,citecolor=blue,filecolor=blue,urlcolor=blue]{hyperref}

\begin{document}

\setlength{\unitlength}{0.0025\columnwidth}

\title{Accelerated FRW Solutions in Chern-Simons Gravity }

\author{Juan Cris\'{o}stomo}
\email{jcrisostomo@udec.cl}
\author{Fernando Gomez}
\email{fernagomez@udec.cl}
\author{Patricio Salgado}
\email{pasalgad@udec.cl}
\affiliation{Departamento de F\'{\i}sica, Universidad de Concepci\'{o}n, Casilla 160-C, Concepci\'{o}n, Chile}

\author{Cristian Quinzacara}
\email{cristian.cortesq@uss.cl}
\affiliation{Departamento de F\'{\i}sica, Universidad de Concepci\'{o}n, Casilla 160-C, Concepci\'{o}n, Chile and\\ Facultad de Ingenier\'ia y Tecnolog\'ia, Universidad San Sebasti\'an, Campus Las Tres Pascualas, Lientur 1457,
Concepci\'{o}n, Chile}

\author{Mauricio Cataldo}
\email{mcataldo@ubiobio.cl}
\affiliation{Departamento de F\'{\i}sica, Universidad del B\'{\i}o-B\'{\i}o, Casilla 5-C,
Concepci\'{o}n, Chile.}

\author{Sergio del Campo}
\email{sdelcamp@ucv.cl}
\affiliation{Instituto de F\'{\i}sica, Pontificia Universidad Cat\'{o}lica de
Valpara\'{\i}so, Av. Universidad 300, Campus Curauma, Valpara\'{\i}so, Chile.}


\date{\today}

\begin{abstract}
We consider a five-dimensional Eins\-tein-Chern-Si\-mons action which is composed of a gravitational sector and a sector of
matter, where the gravitational sector is given by a Chern-Simons gravity
action instead of the Einstein-Hilbert action and where the matter sector is
given by the so called perfect fluid. It is shown that 
\begin{inparaenum}[(i)]
\item the Einstein-Chern-Simons (EChS) field equations subject to suitable conditions can be written in a
similar way to the Einstein-Maxwell field equations;
\item these equations have solutions that describe accelerated expansion for the
three possible cosmological models of the universe, namely, spherical
expansion, flat expansion and hyperbolic expansion when  $\alpha $, a parameter of theory, is
greater than zero. This result allow us to conjeture that this solutions are
compatible with the era of Dark Energy and that the energy-momentum tensor
for the field $h^{a}$, a bosonic gauge field from the Chern-Simons gravity
action, corresponds to a form of positive cosmological
constant.
\end{inparaenum}

It is also shown that the EChS field equations have solutions compatible
with the era of matter:
\begin{inparaenum}[(i)]
\item In the case of an open universe, the solutions correspond to an accelerated expansion ($\alpha >0$) with a minimum scale factor at initial time that, when the time goes to infinity, the scale factor behaves as a hyperbolic sine function. 
\item In the case of a flat universe, the solutions describing an accelerated
expansion whose scale factor behaves as a exponencial function when time
grows. 
\item In the case of a closed universe it is found only one solution
for a universe in expansion, which behaves as a hyperbolic cosine
function when time grows.
\end{inparaenum}
\end{abstract}

\maketitle

\section{Introduction}

Some time ago was shown that the standard, five-di\-men\-sio\-nal General
Relativity can be obtained from Chern-Simons gravity theory for a certain
Lie algebra $\mathfrak{B}$ \cite{Izaurieta2009213}, whose generators $\left\{
\boldsymbol J_{ab},\boldsymbol P_{a},\boldsymbol Z_{ab},\boldsymbol Z_{a}\right\} $ satisfy the commutation relationships

\begin{align*}
\left[\boldsymbol{J}_{ab},\boldsymbol{J}_{cd}\right]\,& =\eta_{ad}\boldsymbol{J}_{bc}-\eta_{ac}\boldsymbol{J}_{bd} +\eta_{bc}\boldsymbol{J}_{ad}-
\eta_{bd}\boldsymbol{J}_{ac},\\
\left[\boldsymbol{P}_{a},\boldsymbol{J}_{bc}\right]\,& =\eta_{ab}\boldsymbol{P}_{c}-\eta_{ac}\boldsymbol{P}_{b},\\
\left[\boldsymbol{J}_{ab},\boldsymbol{Z}_{cd}\right]\,& =\eta_{ad}\boldsymbol{Z}_{bc}-\eta_{ac}\boldsymbol{Z}_{bd} +\eta_{bc}\boldsymbol{Z}_{ad}-
\eta_{bd}\boldsymbol{Z}_{ac},\\
\left[\boldsymbol{Z}_{a},\boldsymbol{J}_{bc}\right]\,& =\eta_{ab}\boldsymbol{Z}_{c}-\eta_{ac}\boldsymbol{Z}_{b},\\
\left[\boldsymbol{P}_{a},\boldsymbol{P}_{b}\right]\,&=\boldsymbol{Z}_{ab},\\
\left[\boldsymbol{P}_{a},\boldsymbol{Z}_{bc}\right]\,& =\eta_{ab}\boldsymbol{Z}_{c}-\eta_{ac}\boldsymbol{Z}_{b}.
\end{align*}

This algebra was obtained from the anti de Sitter (AdS) algebra and a particular semigroup $S$ by means of the S-expansion procedure introduced in Refs.  \cite{izaurieta:123512}, \cite{izaurieta:073511}.

In order to write down a Chern--Simons lagrangian for the $\mathfrak{B}$
algebra, we start from the one-form gauge connection%
\begin{equation}
\boldsymbol{A}=\frac{1}{2}\omega ^{ab}\boldsymbol{J}_{ab}+\frac{1}{l}e^{a}%
\boldsymbol{P}_{a}+\frac{1}{2}k^{ab}\boldsymbol{Z}_{ab}+\frac{1}{l}h^{a}%
\boldsymbol{Z}_{a},
\end{equation}%
and the two-form curvature%
\begin{align}
\boldsymbol{F}&=\frac{1}{2}R^{ab}\boldsymbol{J}_{ab}+\frac{1}{l}T^{a}%
\boldsymbol{P}_{a}+\frac{1}{2}\left( \mathrm{D}_{\omega }k^{ab}+\frac{1}{%
l^{2}}e^{a}e^{b}\right) \boldsymbol{Z}_{ab}\notag\\
& \qquad+\frac{1}{l}\left( \mathrm{D}%
_{\omega }h^{a}+k_{\text{ \ \ }b}^{a}e^{b}\right) \boldsymbol{Z}_{a}.
\end{align}

Consistency with the dual procedure of S-expansion in terms of the Maurer-Cartan forms \cite{izaurieta:073511} demands that $h^{a}$ inherits units of length from the \textsl{f\"{u}nfbein}; that is why it is necessary to
introduce the $l$ parameter again, this time associated with $h^{a}$.

It is interesting to observe that $\boldsymbol{J}_{ab}$ are still Lorentz generators, but $\boldsymbol{P}_{a}$ are no longer AdS boosts; in fact, $\left[ \boldsymbol{P}_{a},\boldsymbol{P}_{b}\right] =\boldsymbol{Z}_{ab}$.
However, $e^{a}$ still transforms as a vector under Lorentz transformations, as it must be in order to recover gravity in this scheme.

A Chern-Simons lagrangian in $d=5$ dimensions is defined to be the following local function of a one-form gauge connection $\boldsymbol{A}$: 
\begin{equation}
L_\text{ChS}^{\left( 5\right) }\left( \boldsymbol{A}\right) =k\left\langle 
\boldsymbol{AF}^{2}-\frac{1}{2}\boldsymbol{A}^{3}\boldsymbol{F+}\frac{1}{10}%
\boldsymbol{A}^{5}\right\rangle ,  \label{lcs}
\end{equation}%
where $\left\langle \cdots \right\rangle $ denotes a invariant tensor for the corresponding Lie algebra$,$ $\boldsymbol F=\textrm d\boldsymbol A+\boldsymbol{AA}$ is the corresponding the two-form curvature and $k$ is a constant \cite{Zanelli:2005sa}.

Using theorem~VII.2 of Ref.~\cite{izaurieta:123512}, it is possible to show that the only
non-vanishing components of a invariant tensor for the \ $\mathfrak{B}$\
algebra are given by%
\begin{eqnarray}
\left\langle \boldsymbol{J}_{a_{1}a_{2}}\boldsymbol{J}_{a_{3}a_{4}}%
\boldsymbol{P}_{a_{5}}\right\rangle &=&\alpha _{1}\frac{4l^{3}}{3}\epsilon
_{a_{1}\cdots a_{5}}, \\
\left\langle \boldsymbol{J}_{a_{1}a_{2}}\boldsymbol{J}_{a_{3}a_{4}}%
\boldsymbol{Z}_{a_{5}}\right\rangle &=&\alpha _{3}\frac{4l^{3}}{3}\epsilon
_{a_{1}\cdots a_{5}},  \notag \\
\left\langle \boldsymbol{J}_{a_{1}a_{2}}\boldsymbol{Z}_{a_{3}a_{4}}%
\boldsymbol{P}_{a_{5}}\right\rangle &=&\alpha _{3}\frac{4l^{3}}{3}\epsilon
_{a_{1}\cdots a_{5}},  \notag
\end{eqnarray}%
where $\alpha _{1}$ and $\alpha _{3}$ are arbitrary independient constants of dimensions $\left[ length\right] ^{-3}$.

Using the extended Cartan's homotopy formula as in Ref. \cite{irs1}, and integrating by parts, it is possible to write down the Chern-Simons Lagrangian in five dimensions for the $\mathcal{B}$ algebra as
\begin{align}
L_\text{EChS}^{(5)}&=\alpha_1l^2\epsilon_{abcde} e^aR^{bc}R^{de}\notag\\
&\quad+\alpha_3\epsilon_{abcde}\left(\frac{2}{3}R^{ab}e^ce^de^e+2l^2k^{ab}R^{cd}T^{e}+l^2R^{ab}R^{cd} h^e\right)\notag\\
&\qquad+\mathrm d B_\text{EChS}^{(4)}\label{1.01}
\end{align}
where the suface term $B_{\text{EChS}}^{(4)}$ is given by
\begin{align}
B_\text{EChS}^{(4)}&=\alpha_1l^2\epsilon_{abcde}e^a\omega^{bc}\left(\frac{2}{3}\mathrm d\omega^{de}+\frac{1}{2}\omega^d_{\phantom 2 f}\omega^{fe}\right)\notag\\
&\quad+ \alpha_3\epsilon_{abcde}\Biggl[l^2\left(h^a\omega^{bc}+k^{ab}e^c\right)\left(\frac{2}{3}\mathrm d\omega^{de}+\frac{1}{2}\omega^d_{\phantom 2 f}\omega^{fe}\right)\notag\\
&\qquad\qquad\qquad+l^2k^{ab}\omega^{cd}\left(\frac{2}{3}\mathrm d e^e+\frac{1}{2}\omega^d_{\phantom 2 f}e^{e}\right)+\frac{1}{6}e^ae^be^c\omega^{de}\Biggr]\label{2.01}
\end{align}
and where $\alpha _{1}$, $\alpha _{3}$ are parameters of the theory, $l$ is a coupling constant, $R^{ab}=\textrm d\omega ^{ab}+\omega^{a}_{\phantom 2 c}\omega^{cb}$ corresponds to
the curvature $2$-form in the first-order formalism related to the $1$-form spin connection \cite{Zanelli:2005sa}, \cite{Chamseddine1989291}, \cite{Chamseddine1990213}, and $e^{a}$, $h^{a}$ and $k^{ab}$ are others gauge fields presents in the theory \cite{Izaurieta2009213}.

From (\ref{1.01}) we can see that the third term is a surface term and can be removed from this Lagrangian. So that,%

\begin{align}
L_\text{EChS}^{(5)}& =\alpha _{1}l^{2}\varepsilon _{abcde}R^{ab}R^{cd}e^{e} 
\notag \\
& \quad +\alpha _{3}\epsilon _{abcde}\left( \frac{2}{3}%
R^{ab}e^{c}e^{d}e^{e}+2l^{2}k^{ab}R^{cd}T^{e}+l^{2}R^{ab}R^{cd}h^{e}\right)
\label{nueve}
\end{align}%
is the Einstein-Chern-Simons Lagrangian studied in Ref \cite{Izaurieta2009213}.

It should be noted the absence of kinetic terms for the fields $h^{a}$ and $k^{ab}$ in equation (\ref{nueve}). The term kinetic for the $h^{a}$ and $k^{ab}$ fields are present in the surface term of the Lagrangian (\ref{1.01})
given by (\ref{2.01}).

The Lagrangian (\ref{nueve}) show that standard, five-dimensional General Relativity emerges as the $l \rightarrow 0$ limit of a CS theory for the generalized Poincar\'{e} algebra $\mathfrak{B}$. Here $l $ is a length
scale, a coupling constant that characterizes different regimes within the
theory. The $\mathfrak{B}$\ algebra, on the other hand, is constructed from the AdS algebra and a particular semigroup $S$ by means of the S-expansion procedure. The field content induced by the $\mathfrak{B}$ algebra includes the vielbein $e^{a}$, the spin connection $\omega ^{ab}$ and two extra bosonic fields $h^{a}$ and $k^{ab}$, which can be interpreted
as boson fields coupled to the field curvature and the parameter $l^{2}$ can be interpreted as a kind of coupling constant.

Recently was found \cite{PhysRevD.84.063506} that the standard five-di\-men\-sio\-nal FRW
equations and some of their solutions can be obtained, in a certain limit,
from the so-called Chern-Simons-FRW field equations, which are the
cosmological field equations corresponding to a Chern-Simons gravity theory.

It is the purpose of this paper to show that the Einstein-Chern-Simons
(EChS) field equations, subject to
\begin{inparaenum}[(i)]
 \item the tor\-sion-free condition ($T^{a}=0$) and 
 \item the variation of the matter Lagrangian with respect to (\textit{w.r.t.}) the spin connection is zero ($\delta L_{M}/\delta\omega ^{ab}=0$) can be written in a similar way to the Einstein-Maxwell field equations. 
 \end{inparaenum}
The interpretation of the $h^{a}$ field as a perfect fluid\- allow\- us to show that the
Einstein-Chern-Simons field equations have an universe in accelerated
expansion as a of their solutions.

This paper is organized as follows: In Section \ref{sec02} we brie\-fly review
the Einstein-Chern-Simons field equations. In Sec\-tion \ref{sec030} we study the Einstein-Chern-Simons field
equations in the range of validity of general relativity. In Section \ref{sec03} we consider accelerated solutions for Eins\-tein-Chern-Si\-mons field equations. We try to find solutions that describes accelerated expansion for
cases of open universes, flat universes and closed universes. In Section \ref{sec04} we consider the consistency of the solutions with the "Era of Matter". A summary and an appendix conclude this work.

\section{Einstein-Chern-Simons field equations\label{sec02}}


In Ref. \cite{PhysRevD.84.063506} was found that in the presence of matter the lagrangian is
given by 
\begin{equation}
L=L_{\mathrm{ChS}}^{(5)}+\kappa L_{M}  \label{1a}
\end{equation}
where $L_{\mathrm{ChS}}^{(5)}$ is the five-dimensional Chern-Simons la\-gran\-gian given by (\ref{nueve}),\linebreak $L_{M}=L_{M}(e^{a},h^{a},\omega^{ab})$ is the matter Lagrangian and $\kappa$ is a coupling constant related to the effective Newton's constant. The variation of the lagrangian (\ref{1a}) w.r.t. the dynamical fields vielbein $e^a$, spin connection $\omega^{ab}$, $h^a$ and $k^{ab}$, leads to the following field equations

\begin{align}
\varepsilon _{abcde}\Bigl( 2\alpha _{3}R^{ab}e^{c}e^{d}+\alpha
_{1}l^{2}R^{ab}R^{cd}\qquad\quad&\notag\\
+2\alpha _{3}l^{2}D_{\omega }k^{ab}R^{cd}\Bigr)
&=\kappa \frac{\delta L_{M}}{\delta e^{e}},  \label{2}
\end{align}
\begin{equation}
\alpha _{3}l^{2}\varepsilon _{abcde}R^{ab}R^{cd}=\kappa \frac{\delta L_{M}}{%
\delta h^{e}},  \label{3}
\end{equation}%
\begin{equation}
2\alpha _{3}l^{2}\varepsilon _{abcde}R^{cd}T^{e}=\kappa \frac{\delta L_{M}}{%
\delta k^{ab}} , \label{4}
\end{equation}%
\begin{align}
2\varepsilon _{abcde}\Bigl( \alpha _{1}l^{2}R^{cd}T^{\text{ }e}+\alpha
_{3}l^{2}D_{\omega }k^{ab}T^{e}\quad&\notag\\
+\alpha _{3}e^{c}e^{d}T^{e}+\alpha
_{3}l^{2}R^{cd}D_{\omega }h^{e}\Bigr)\ &   \notag \\
+2\alpha _{3}\varepsilon _{abcde}l^{2}R^{cd}k_{\text{ }f}^{e}e^{f} &=\kappa 
\frac{\delta L_{M}}{\delta \omega ^{ab}} . \label{5}
\end{align}

For simplicity, we will assume that the torsion vanishes ($T^{a}=0$) and $k^{ab}=0$. In this case the Eqs.(\ref{2} - \ref{5}) takes the form

\begin{align}
\varepsilon_{abcde}\left(
2\alpha_{3}R^{ab}e^{c}e^{d}+\alpha_{1}l^{2}R^{ab}R^{cd}\right)& =\kappa\frac{%
\delta L_{M}}{\delta e^{e}},\label{3p}\\
\alpha_{3}l^{2}\varepsilon_{abcde}R^{ab}R^{cd}&=\kappa\frac{\delta L_{M}}{%
\delta h^{e}},\label{4p}\\
\frac{\delta L_{M}}{\delta k^{ab}}&=0\label{5p}\\
2\alpha_{3}l^{2}\varepsilon_{abcde}R^{cd}D_{\omega}h^{e}&=\kappa\frac{\delta
L_{M}}{\delta\omega^{ab}}\label{6p}.
\end{align}

This field equations system can be written in the form%
\begin{align}
\varepsilon_{abcde}R^{ab}e^{c}e^{d}&=4\kappa_{5}\left( \frac{\delta L_{M}}{%
\delta e^{e}}+\alpha\frac{\delta L_{M}}{\delta h^{e}}\right) , \label{6}\\
l^{2}\varepsilon_{abcde}R^{ab}R^{cd}&=8\kappa_{5}\frac{\delta L_{M}}{\delta
h^{e}} , \label{7}\\
l^{2}\varepsilon_{abcde}R^{cd}D_{\omega}h^{e}&=4\kappa_{5}\frac{\delta L_{M}}{%
\delta\omega^{ab}}  \label{8}
\end{align}
where we introduce $\kappa_{5}=\kappa/8\alpha_{3}$ and $\alpha=-\alpha_{1}/\alpha_{3}$.

The field equation (\ref{2}) contains three terms. The first one,
proportional to the Einstein tensor. The second one corresponds to a quadratic
term in the curvature, and a third one, a term that describes the dynamics of the
field $k^{ab}$. Since we asume $k^{ab}=0$ the last term in left side of Eq. (\ref{2}) vanishes.

In order to write this field equation manner analogous to Einstein's
equations, one chooses to leave the term proportional to the Einstein tensor
on the left side of Eq. (\ref{2}) 
\begin{equation*}
\epsilon _{abcde}R^{bc}e^{d}e^{e}=\frac{\kappa }{2\alpha _{3}}\frac{\delta
L_{M}}{\delta e^{a}}-\frac{\alpha _{1}}{2\alpha _{3}}l^{2}\,\epsilon
_{abcde}R^{bc}R^{de}
\end{equation*}%
and using the Eq. (\ref{4p}) we obtain Eq. (\ref{6}).

This result allows us to interpret $\delta L_{M}/\delta h^{a}$ as
the energy momentum tensor for a second type of matter, not ordinary. Henceforth we will say that $\delta L_{M}/\delta h^{a}$ corresponds to the energy-momentum tensor for the field $h^{a}$. 

The equation of motion for the $h^{a}$-field is given by Eq.(\ref{8}). The
condition $\delta L_{M}/\delta \omega ^{ab}=0$ (usual in gravity
theories), imposed for consistency with the condition $T^{a}=0$, leads to the
equation of motion (\ref{11}) for the $h^{a}$-field . This means that $h^{a}$-field is governed by the following field equations

\begin{align}
\varepsilon_{abcde}R^{ab}e^{c}e^{d}&=4\kappa_{5}\left( \frac{\delta L_{M}}{%
\delta e^{e}}+\alpha\frac{\delta L_{M}}{\delta h^{e}}\right),  \label{9}\\
\frac{l^{2}}{8\kappa_{5}}\varepsilon
_{abcde}R^{ab}R^{cd}&=\frac{\delta L_{M}}{\delta h^{e}},  \label{10}\\
\varepsilon_{abcde}R^{cd}D_{\omega}h^{e}&=0.  \label{11}
\end{align}

This means that the Einstein-Chern-Simons field equations, subject to the
conditions $T^{a}=0$,\ $k^{ab}=0$ and $\delta L_{M}/\delta\omega^{ab}=0$,
can be re-written in a way similar to the Einstein-Maxwell field equations.

From (\ref{9}-\ref{11}) we can see that if $L_{M}=0$, then in five
dimensions there is no solution of  Schwarzschild type \cite{Izaurieta2009213}, \cite{PhysRevD.85.124026}.

\section{Einstein-Chern-Simons Equations in the range of validity
of General Relativity\label{sec030}}

From (\ref{9}-\ref{10}) we can see that general relativity is valid when \begin{inparaenum}[(i)]
\item the curvature $R^{ab}$ takes values not excessively large
\item the parameter $l$ takes small values $\left( l\longrightarrow 0\right) $ \cite{Izaurieta2009213};
\item the constant $\alpha $ takes values not excessively large.
\end{inparaenum} 
In fact, in this case we have that (\ref{10}) takes the form
\begin{equation}
\frac{\delta L_{M}}{\delta h^{e}}\approx 0 . \label{13'}
\end{equation}

Introducing (\ref{13'}) into (\ref{9}) we obtain the Einstein's field
equation

\begin{equation}
\varepsilon _{abcde}R^{ab}e^{c}e^{d}\approx 4\kappa _{5}\frac{\delta L_{M}}{%
\delta e^{e}}.  \label{14}
\end{equation}

If $R^{ab}$ is not large then $\delta L_{M}/\delta e^{a}$ is also not large.
This means that General Relativity can be seen as a low energy limit of
Einstein-Chern-Simons gravity. So that, in the range of validity of the
General Relativity, the equations (\ref{9}-\ref{11}) are given by 
\begin{align}
\varepsilon _{abcde}R^{ab}e^{c}e^{d}&=4\kappa _{5}\frac{\delta L_{M}}{\delta
e^{e}},  \label{15}\\
\varepsilon _{abcde}R^{cd}D_{\omega }h^{e}&=0.  \label{16}
\end{align}

On the another hand, if $R^{ab}$ is large enough, so that when it is
multiplied by $l^{2}$ (which is very small) will have a non-negligible
results, then we will find that $\delta L_{M}/\delta h^{a}$ is not
negligible. \ This means that, in this case, we must consider the entire
system of equations (\ref{9}-\ref{11}).

\section{Einstein-Chern-Simons Field Equa\-tions for a Friedmann-Robertson-Walker-like spacetime\label{sec025}}

The shape of the field $e^{a}$ is obtained from of the application of the
cosmological principle to the metric tensor of spacetime: it is considered a splitting of the 5D-manifold in a maximally symmetric four-dimensional
manifold  and one temporal dimension ($M=R\times \Sigma _{4}$). This leads to five dimensional Friedmann-Robertson-Walker (FRW) metric. So that, the vielbein can be chosen like in \cite{PhysRevD.84.063506}:

\begin{align}
e^{0}& \displaystyle =dt,  \notag \\
e^{1}& \displaystyle =\frac{a(t)}{\sqrt{1-kr^{2}}}\,dr  ,\notag \\
e^{2}& \displaystyle =a(t)r\,d\theta _{2} , \notag \\
e^{3}& \displaystyle =a(t)r\sin\theta _{2}\,d\theta _{3} , \notag \\
e^{4}& \displaystyle =a(t)r\sin\theta _{2}\sin\theta _{3}\,d\theta _{4}  \label{17}
\end{align}
where $a(t)$ is the scale factor of the universe and $k$ is the sign of the curvature of space ($\Sigma_4$): 
\begin{inparaenum}[(i)]
\item $+1$ for a closed space ($S^4$),
\item $0$ for a flat space ($E^4$) and
\item $-1$ for an open space (hyperbolic).
\end{inparaenum}

The application of the cosmological principle to the metric tensor of the spacetime also constrains the shape of the field $h^a$ (see for example \cite{PhysRevD.84.063506}). A detailed discussion can be also found in Ref. \cite{weinberg1972gravitation}. The bosonic field $h^{a}$ is given by
\begin{align}
h^{0}& =h(0)\,dt=h(0)e^{0},  \notag \\
h^{1}& =h(t)\frac{a(t)}{\sqrt{1-kr^{2}}}\,dr=h(t)e^{1} , \notag \\
h^{2}& =h(t)a(t)r\,d\theta _{2}=h(t)e^{2} , \notag \\
h^{3}& =h(t)a(t)r\sin\theta _{2}\,d\theta _{3}=h(t)e^{3}  ,\notag \\
h^{4}& =h(t)a(t)r\sin\theta _{2}\sin\theta _{3} \,d\theta _{4}=h(t)e^{4}
\label{18}
\end{align}%
where $h(0)$ is a constant and $h(t)$ is a function of time $t$ that must be determined. Substituting (\ref{18}) into Eq. (\ref{11}) we obtain the explicit form of the equations of motion for the $h^{a}$-field, which will be displayed in Eq.(\ref{eqz05}).

In accordance with the equation (\ref{9}), we will consider a fluid composed of two perfect fluids, the first one related to ordinary energy-momentum tensor ($T_{\mu\nu}\sim \frac{\delta L_{M}}{\delta e^{a}}$) and the second one related to field $h^a$  ($T_{\mu\nu}^{(h)}\sim\frac{\delta L_{M}}{\delta h^{a}} $). The energy-momentum
tensors in the comoving frame, are given by
\begin{equation}
T_{\mu\nu}=\textrm{diag}(\rho,p,p,p,p) , \label{19}
\end{equation}%
\begin{equation}
T_{\mu\nu}^{(h)}=\textrm{diag}\Bigl(\rho^{(h)},p^{(h)},p^{(h)},p^{(h)},p^{(h)}\Bigr) , \label{20}
\end{equation}
where $\rho$ is the matter density and $p$ is the pressure of fluid. Then, the energy-momentum tensor for the composed fluid is 
\begin{align}
\tilde{T}_{\mu\nu} & =T_{\mu\nu}+\alpha T_{\mu\nu}^{(h)}  \label{21} \\
& =\textrm{diag}\Bigl(\rho+\alpha\rho^{(h)},p+\alpha p^{(h)},\notag\\
&\qquad\qquad\quad p+\alpha p^{(h)},p+\alpha p^{(h)},p+\alpha p^{(h)}\Bigr)  \label{22} \\
& =\textrm{diag}(\tilde{\rho},\tilde{p},\tilde{p},\tilde{p},\tilde{p}).  \label{23}
\end{align}

In the torsion-free case, the energy momentum tensor of ordinary
matter satisfies a conservation equation and the Einstein tensor has also zero
divergence. In this case the energy momentum tensor for the non-ordinary matter must
also satisfy a conservation equation. In fact, from Eq. (\ref{9}) we find

\begin{equation}\label{ect01}
\nabla_\mu T^{\mu}_{\phantom b \nu}=0\quad,\quad \nabla_\mu T^{(h)\,\mu}_{\phantom{bbbbb} \nu}=0
\end{equation}

Introducing (\ref{17} - \ref{23}) into eqs. (\ref{9} - \ref{11}) we find the following field equations (see Ref. \cite{PhysRevD.84.063506} and Appendix \ref{apendixa})
\begin{align}
6\left( \frac{{\dot a}^{2}+k}{a^{2}}\right)^{\phantom{b}}&=\kappa _{5}\tilde{\rho}  \label{eqz01},\\
3\left[ \frac{{\ddot a}}{a}+\left( \frac{{\dot a}^{2}+k}{a^{2}}\right) \right]^{\phantom{b}}& =-\kappa _{5}\tilde{p}  \label{eqz02},\\
{\frac{3l^{2}}{\kappa _{5}}\left( \frac{{\dot a}^{2}+k}{a^{2}}\right) ^{2}}^{\phantom{b}}&=\rho ^{(h)}  \label{eqz03},\\
\frac{3l^{2}}{\kappa _{5}}\frac{{\ddot a}}{a}\left( \frac{{\dot a}^{2}+k}{a^{2}}\right) ^{\phantom{b}}&=-p^{(h)}  \label{eqz04},\\
\left( \frac{{\dot a}^{2}+k}{a^{2}}\right) \left[ \left(h-h(0)\right) \frac{{\dot a}}{a}+{\dot h}\right]^{\phantom{b}} &=0
\label{eqz05}.
\end{align}

We should note that equation (\ref{eqz01}) was studied in Ref. \cite{1475-7516-2012-12-005} in the context of
inflationary cosmology .


The Equations (\ref{eqz01}) and (\ref{eqz02}) are very similar to the
Friedmann equations in five dimensions. However now $\rho$ and $p$ are
subject to restrictions imposed by the remaining equations.

\section{Acelerated Solution for Einstein-Chern-Simons Field Equa\-tions\label{sec03}}

In order to recover the known results of the standard cosmology in the
context of accelerated expansion we use the approach 
\begin{equation*}
T_{\mu \nu }\ll \alpha T_{\mu \nu }^{(h)}
\end{equation*}%

This approach is analogous to the case when, in the era of Dark Energy, the energy momentum tensor is neglected
compared to the cosmological constant. This means that the contribution from the ordinary matter is negligible compared to the contribution from the field $h^a$. In this case, the energy-momentum tensor $\tilde T_{\mu\nu}$ fluid is given by
\begin{align}
\tilde T_{\mu\nu} & =\text{diag}(\tilde\rho, \tilde p, \tilde p, \tilde p,
\tilde p) \notag\\
& =\alpha T_{\mu\nu}^{(h)}\notag\\
&=\text{diag}\Bigl(\alpha\rho^{(h)} ,\alpha p^{(h)}
,\alpha p^{(h)} , \alpha p^{(h)} ,\alpha p^{(h)}\Bigr)
\end{align}
and the equations (\ref{eqz01} - \ref{eqz05}) take the form
\begin{align}  
6\left(\frac{{\dot a}^{2}+k}{a^{2}}\right)^{\phantom{b}} &=\kappa_{5}\alpha\rho^{(h)},\label{eqz06}\\
 3\left[ \frac{\ddot a}{a}+\left( \frac{{\dot a}^{2}+k}{a^{2}}\right) \right]^{\phantom{b}}&=-\kappa_{5}\alpha p^{(h)},\label{eqz07}\\
{\frac{3l^{2}}{\kappa_{5}}\left( \frac{{\dot a}^{2}+k}{a^{2}}\right)^{2}}^{\phantom{b}}&=\rho^{(h)},\label{eqz08}\\
\frac{3l^{2}}{\kappa_{5}}\frac{\ddot a}{a}\left( \frac{{\dot a}^{2}+k}{a^{2}}\right)^{\phantom{b}} &=-p^{(h)},\label{eqz09}\\
\left( \frac{{\dot a}^{2}+k}{a^{2}}\right) \left[ (h-h(0))\frac{\dot a}{a}+\dot h\right]^{\phantom{b}} &=0.\label{eqz10}
\end{align}

\subsection{Case $T_{\protect\mu \protect\nu }=0$ and $k=-1$}

Introducing (\ref{eqz08}) into (\ref{eqz06}) we obtain
\begin{equation}
6\left( \frac{{\dot{a}}^{2}+k}{a^{2}}\right) =3l^{2}\alpha\left( \frac{{\dot{%
a}}^{2}+k}{a^{2}}\right) ^{2}
\end{equation}
which can be rewritten

\begin{equation}  \label{eqz11}
\left( \frac{{\dot a}^{2}+k}{a^{2}}\right) \left( \frac{2}{\alpha l^{2}}- 
\frac{{\dot a}^{2}+k}{a^{2}}\right) =0.
\end{equation}

\subsubsection{Solution $\ddot{a}=0$}

Consider the solution $\ddot{a}=0$, i.e., a solution without
accelerated expansion. For the first term in left side of (\ref{eqz11}) we
have 
\begin{equation}
\frac{{\dot{a}}^{2}+k}{a^{2}}=0,
\end{equation}%
remembering $k=-1,$ we have 
\begin{equation}
\dot{a}=\sqrt{-k}.
\end{equation}%
The solution is%
\begin{equation}
a(t)=\sqrt{-k}\left( t-t_{0}\right) +a_{0}.
\end{equation}%
In this case $a(t)$ is increase linearly, i.e., there is no accelerated
expansion.

\begin{figure}[h]
\includegraphics[width=350\unitlength]{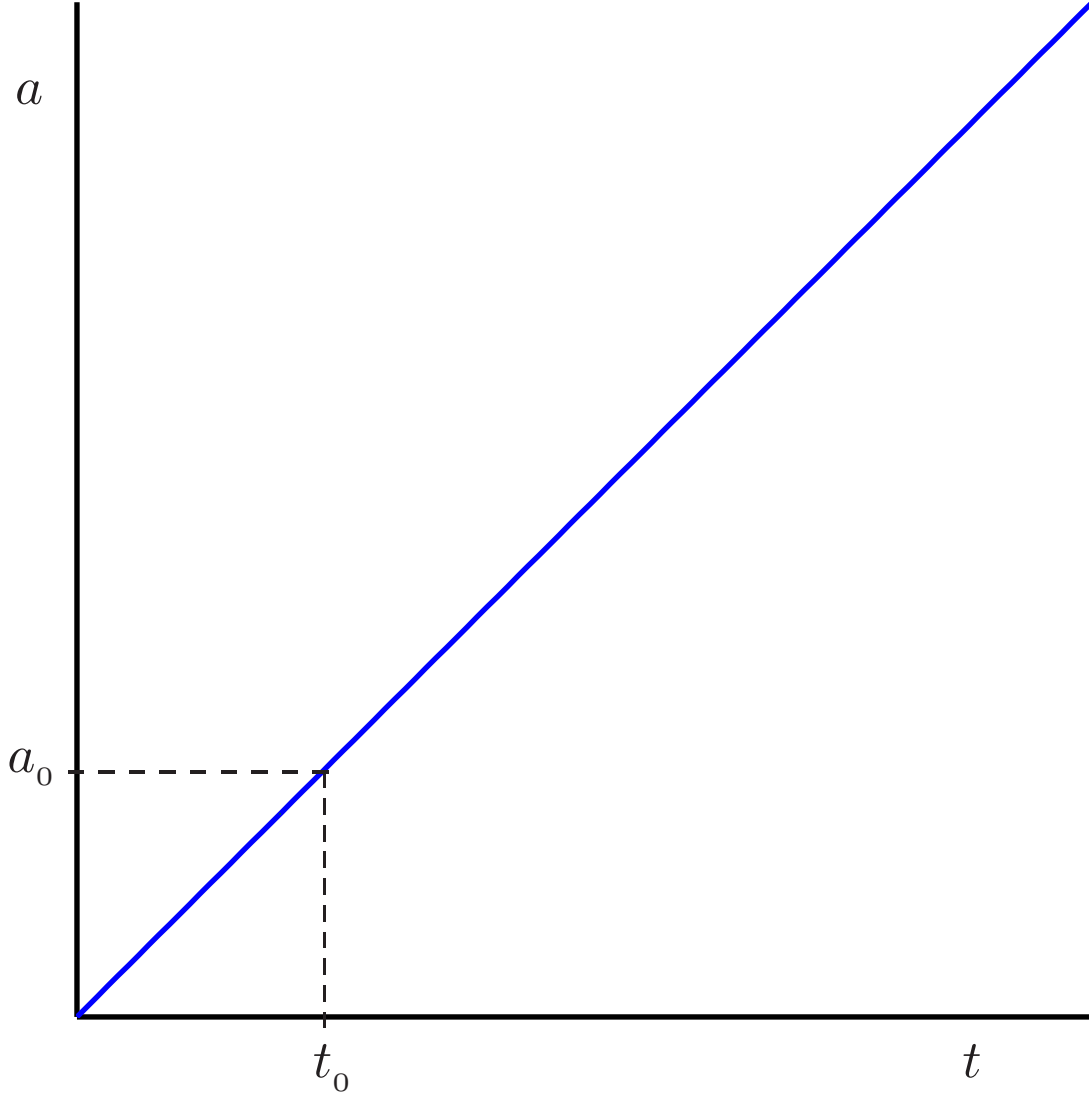}
\centering
\caption{Graph of $a(t)=\sqrt{-k}\,(t-t_0)+a_0$ ($k=-1$).}
\end{figure}

Replacing this solutions into equations (\ref{eqz06} - \ref{eqz09}) we find%
\begin{equation}
\rho ^{(h)}=p^{(h)}=0
\end{equation}%
and equation (\ref{eqz10}) is satisfied for  $h(t)$ arbitrary.

\subsubsection{Solution $\ddot{a}\neq 0 $}

From (\ref{eqz11}) we obtain we obtain 
\begin{equation}
{\dot{a}}^{2}-\frac{2}{\alpha l^{2}}a^{2}=-k . \label{eqz12}
\end{equation}
From (\ref{eqz12}) we can see two options 
\begin{inparaenum}[(i)]
\item $\alpha >0$ and 
\item $\alpha <0$.
\end{inparaenum}
\newline

\paragraph{Case $\protect\alpha >0:$}\ 

\begin{figure}[h]
\includegraphics[width=350\unitlength]{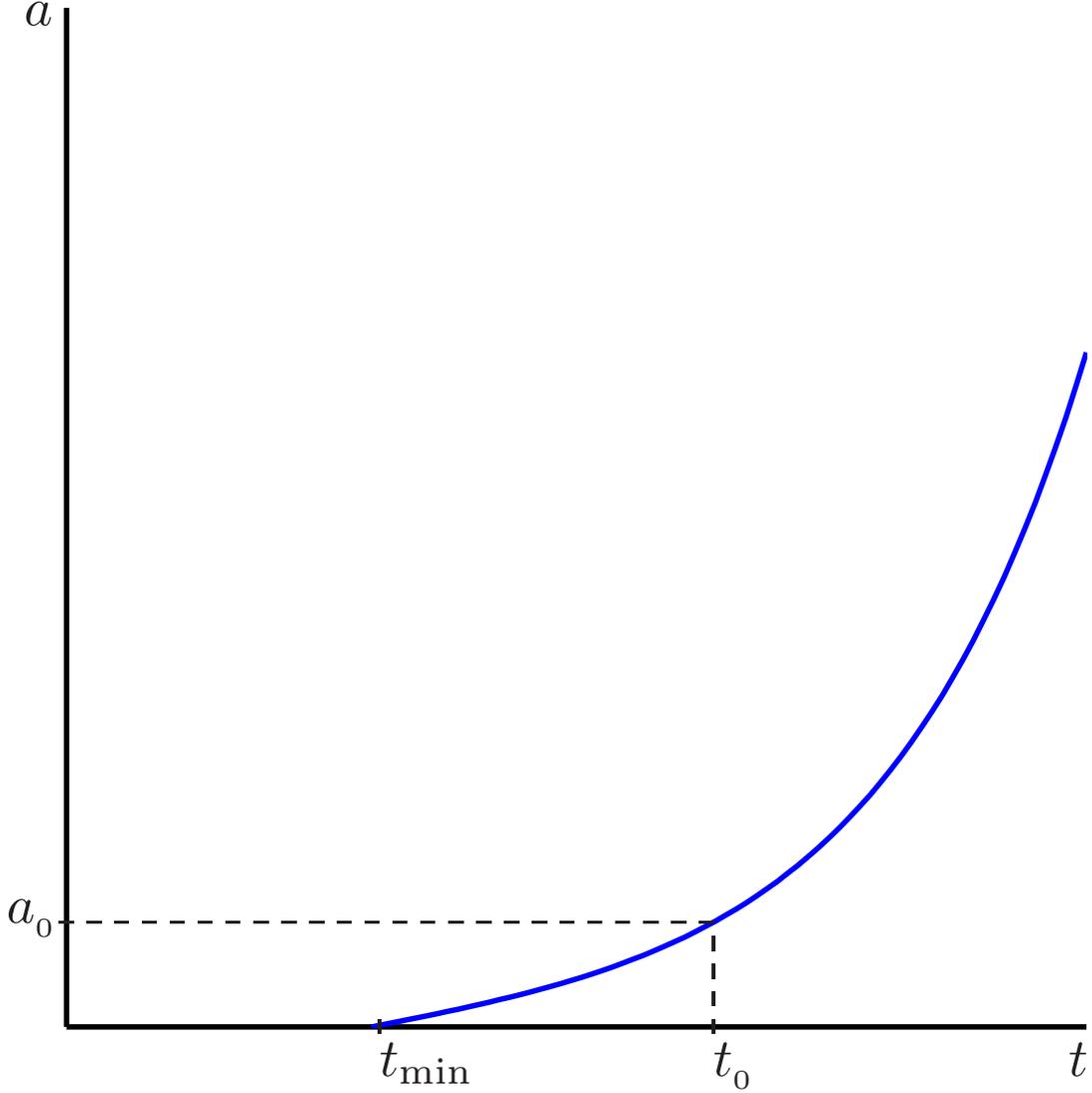}
\centering
\caption{ Graph of $a(t)$ with $\alpha>0$ and $k=-1$. See equation (\ref{eqz13}).\label{fig02}}
\end{figure}

Consider the case where the constant $\alpha$ is positive. Using the following 
\textit{ansatz}\footnote{This \textit{ansatz} can be obtained from 
\begin{equation*}
\dot{a}=\sqrt{\frac{2}{\alpha l^{2}}a^{2}-k}
\end{equation*}%
whose solution is ($\alpha >0$, $k=-1$) 
\[
\protect\int_{t'}^t\frac{da}{\sqrt{\frac{2}{\alpha l^{2}}a^{2}-k}}=t-t^{\prime }
\]
using an hyperbolic substitution 
\begin{equation*}
\sqrt{\frac{\alpha l^{2}}{2}}\text{arsinh}\left( \sqrt{-\frac{2}{\alpha
l^{2}k}}\,a\right) =t-t^{\prime }.
\end{equation*}}

\begin{equation}
a(t)=A\sinh \left( \sqrt{\frac{2}{\alpha l^{2}}}(t-t^{\prime })\right) 
\end{equation}%
where $t^{\prime }$ is a constant of integration, we obtain 
\begin{equation}
A=\sqrt{-\frac{\alpha l^{2}k}{2}}
\end{equation}%
and therefore 
\begin{equation}
a(t)=\sqrt{-\frac{\alpha l^{2}k}{2}}\sinh \left( \sqrt{\frac{2}{\alpha l^{2}}
}(t-t^{\prime })\right) ,
\end{equation}%
the initial condition $a_{0}=a(t=t_{0})$ leads 
\begin{align}
a(t)&=\sqrt{-\frac{\alpha l^{2}k}{2}}\notag\\
 &\quad\ \times\sinh\Biggl[\sqrt{\frac{2}{\alpha l^{2}}%
}(t-t_{0})+\textrm{arsinh} \left( \sqrt{-\frac{2}{\alpha l^{2}k}}\,a_{0}\right) %
\Biggr]  \label{eqz13}
\end{align}%
and 
\begin{align}
\dot{a}(t)&=\sqrt{-k}\notag\\
&\quad \times\cosh \left[ \sqrt{\frac{2}{\alpha l^{2}}}%
(t-t_{0})+\textrm{arsinh} \left( \sqrt{-\frac{2}{\alpha l^{2}k}}\,a_{0}\right) %
\right]  . \label{eqz14}
\end{align}
This results shows that if\textbf{\ }$\alpha >0$, then there is an
accelerated expansion (see Fig. \ref{fig02}).

On the another hand, from (\ref{eqz13}) and (\ref{eqz14}) we can see that 
\begin{equation}\label{eqz14.5}
\ddot{a}(t)=\frac{2}{\alpha l^{2}}a(t),
\end{equation}%
replacing (\ref{eqz13}), (\ref{eqz14}) and (\ref{eqz14.5}) into (\ref{eqz06} - \ref{eqz09})
we obtain%
\begin{equation}
\rho ^{(h)}=-p^{(h)}=\frac{12}{\kappa _{5}\alpha l^{2}},
\end{equation}%
i.e., we have an accelerated expansion when the energy density is positive
and pressure is negative (like a cosmological constant positive).

From equation (\ref{eqz10}) we find

\begin{equation}
-\frac{\dot{h}}{h-h(0)}=\frac{\dot{a}}{a}.
\end{equation}%

Integrating, we find 
\begin{equation}
h(t)=\frac{C}{\sinh \left[ \sqrt{\frac{2}{\alpha l^{2}}}(t-t_{0})+\text{%
arsinh}\left( \sqrt{-\frac{2}{\alpha l^{2}k}}\,a_{0}\right) \right] }+h(0)
\end{equation}%
where $C$ is a constant of integration. The initial condition $%
h_{0}=h(t_{0}) $ leads
\begin{equation*}
h(t)=\frac{\Bigl(h_{0}-h(0)\Bigr)\sqrt{\frac{2}{\alpha l^{2}k}}\,a_{0}}{\sinh \left[ 
\sqrt{\frac{2}{\alpha l^{2}}}(t-t_{0})+\text{arsinh}\left( \sqrt{-\frac{2}{%
\alpha l^{2}k}}\,a_{0}\right) \right] }+h(0)
\end{equation*}%
from where we can see that $h(t)\rightarrow h(0)$ when $t\rightarrow \infty $
\newline

\paragraph{Case $\protect\alpha <0:$}\ 

Consider now the case when the constant $\alpha $ is negative. The \textit{ansatz}

\begin{equation}
a(t)=A\sin \left( \sqrt{-\frac{2}{\alpha l^{2}}}(t-t^{\prime })\right)
\end{equation}%
with $t^{\prime }$ a contant of integration, leads 
\begin{equation}
A=\sqrt{\frac{\alpha l^{2}k}{2}},
\end{equation}%
therefore 
\begin{equation}
a(t)=\sqrt{\frac{\alpha l^{2}k}{2}}\sin \left( \sqrt{-\frac{2}{\alpha l^{2}}}%
(t-t^{\prime })\right).
\end{equation}%
The initial condition $a_{0}=a(t=t_{0}),$ leads 
\begin{align}
a(t)&=\sqrt{\frac{\alpha l^{2}k}{2}}\notag\\
&\quad\ \times\sin \left[ \sqrt{-\frac{2}{\alpha l^{2}}}%
(t-t_{0})+\arcsin \left( \sqrt{\frac{2}{\alpha l^{2}k}}\,a_{0}\right) \right]
\label{eqz15}
\end{align}%
and%
\begin{align}
\dot{a}(t)&=\sqrt{-k}\notag\\
&\quad\times \cos \left[ \sqrt{-\frac{2}{\alpha l^{2}}}%
(t-t_{0})+\arcsin \left( \sqrt{\frac{2}{\alpha l^{2}k}}\,a_{0}\right) \right].
\label{eqz16}
\end{align}

\begin{figure}[h]
\includegraphics[width=350\unitlength]{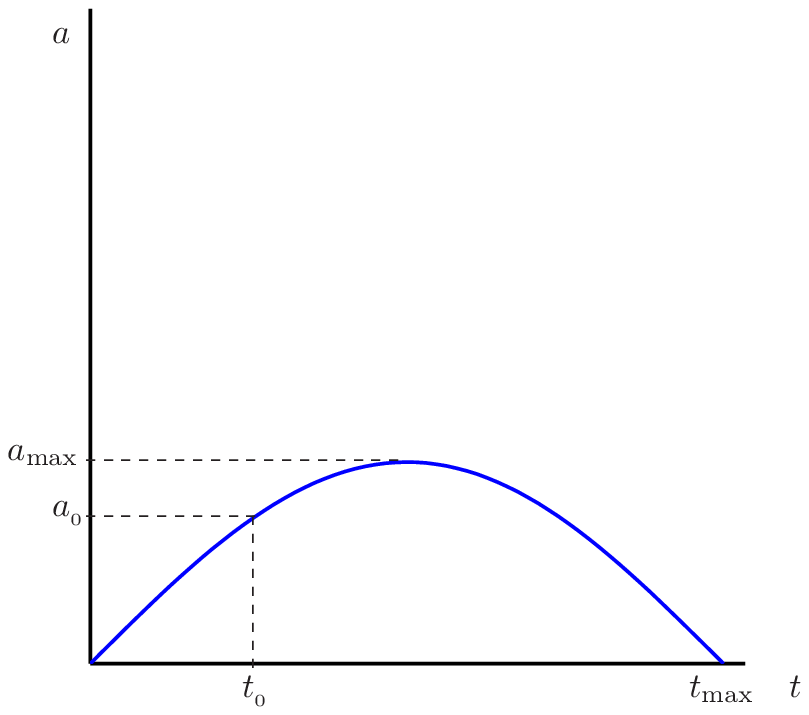}
\centering
\caption{ Graph of $a(t)$ with $\alpha<0$ and $k=-1$. See equation (\ref{eqz15}).\label{fig03}}
\end{figure}

Therefore if $a(t)>0$ then $\ddot{a}(t)<0$, which shows that if $\alpha <0$, then there is a decelerated expansion (see Fig. \ref{fig03}).

On the another hand, replacing (\ref{eqz15}) and (\ref{eqz16}) into (\ref{eqz06} - \ref{eqz09})
we obtain%
\begin{equation}
\rho ^{(h)}=-p^{(h)}=\frac{12}{\kappa _{5}\alpha l^{2}}.
\end{equation}

Since the energy momentum tensor is given by
\begin{equation}
\tilde{T}_{\mu \nu }=\alpha T_{\mu \nu }^{(h)}=\text{diag}\Bigl(\alpha \rho
^{(h)},\alpha p^{(h)},\alpha p^{(h)},\alpha p^{(h)},\alpha p^{(h)}\Bigr)
\end{equation}%
we have that the corresponding energy density and pressure are ($\alpha <0$)%
\begin{equation}
\tilde{\rho}=\alpha \rho ^{(h)}=\frac{12}{\kappa _{5}\alpha l^{2}}<0,
\end{equation}%
\begin{equation}
\tilde{p}=\alpha p^{(h)}=-\frac{12}{\kappa _{5}\alpha l^{2}}>0,
\end{equation}%
i.e., the energy density is negative and the pressure is positive (like a
cosmological constant negative)\textbf{.}

From equation (\ref{eqz10}) we find
\begin{equation}
-\frac{\dot{h}}{h-h(0)}=\frac{\dot{a}}{a}.
\end{equation}%
Integrating, we find 
\begin{equation}
h(t)=\frac{C}{\sin \left[ \sqrt{-\frac{2}{\alpha l^{2}}}(t-t_{0})+\text{%
arcsin}\left( \sqrt{\frac{2}{\alpha l^{2}k}}\,a_{0}\right) \right] }+h(0)
\end{equation}%
where $C$ is a constant of integration. The initial condition $%
h_{0}=h(t_{0}) $, leads
\begin{equation}
h(t)=\frac{(h_{0}-h(0))\sqrt{\frac{2}{\alpha l^{2}k}}\,a_{0}}{\sin \left[ 
\sqrt{-\frac{2}{\alpha l^{2}}}(t-t_{0})+\text{arcsin}\left( \sqrt{\frac{2}{%
\alpha l^{2}k}}\,a_{0}\right) \right] }+h(0).
\end{equation}

\subsection{Case $T_{\protect\mu \protect\nu }=0$ and $
k=0 $}

Introducing (\ref{eqz08}) into (\ref{eqz06}) and considering $k=0$, we obtain

\begin{equation}
6\left( \frac{\dot{a}}{a}\right) ^{2}=3l^{2}\alpha \left( \frac{\dot{a}}{a}%
\right) ^{4}
\end{equation}%
which can be rewritten as

\begin{equation}  \label{eqz17}
\left( \frac{\dot a}{a}\right) ^{2}\left( \frac{2}{\alpha l^{2}}- \frac{{%
\dot a}^{2}}{a^{2}}\right) =0.
\end{equation}

\subsubsection{Static solution $\dot{a}=0$}

The solution for an static universe is given by%
\begin{equation}
a(t)=a_{0}
\end{equation}%
which leads 
\begin{equation}
\rho ^{(h)}=p^{(h)}=0
\end{equation}%
and the equation (\ref{eqz10}) is satisfied for all $h(t)$.

\subsubsection{Non-static solution $\dot{a}\neq 0$}

From (\ref{eqz17}) we obtain 
\begin{equation}
{\dot{a}}^{2}-\frac{2}{\alpha l^{2}}a^{2}=0.  \label{eqz18}
\end{equation}%
This equation have solution, \textit{only if} $\alpha >0$.
\newline

\paragraph{Case $\protect\alpha >0:$}\ 

In this case we have an expanding universe
\begin{equation}
a(t)=A\exp \left( \sqrt{\frac{2}{\alpha l^{2}}}\,t\right) .
\end{equation}

The initial condition $a_{0}=a(t_{0})$ leads%
\begin{equation}
a(t)=a_{0}\exp \left( \sqrt{\frac{2}{\alpha l^{2}}}\,\left( t-t_{0}\right)
\right)  \label{100}
\end{equation}%
and%
\begin{equation*}
\rho ^{(h)}=-p^{(h)}=\frac{12}{\kappa _{5}\alpha l^{2}}.
\end{equation*}

\begin{figure}[h]
\includegraphics[width=350\unitlength]{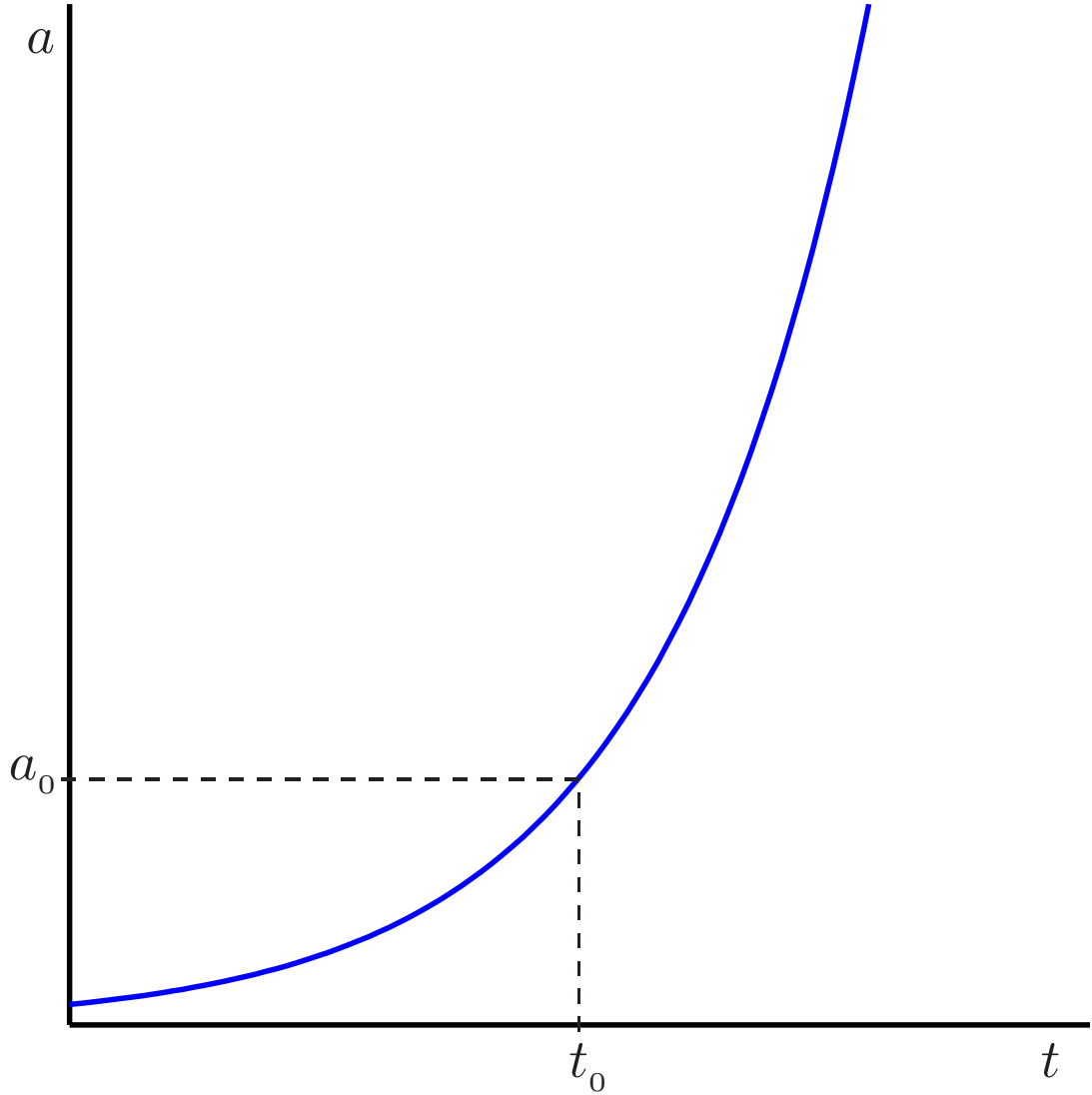}
\centering
\caption{ Graph of $a(t)=a_0\exp\left(\sqrt{\frac{2}{\alpha l^2}}(t-t_0)\right)$}
\end{figure}

Replacing (\ref{100}) into equation (\ref{eqz10}), solving for $h(t)$ and using the initial condition $h_{0}=h(t_{0})$, we find%
\begin{equation}
h(t)=\frac{h_{0}-h(0)}{\exp \left( \sqrt{\frac{2}{\alpha l^{2}}}%
(t-t_{0})\right) }+h(0).
\end{equation}
\newline

\paragraph{Case $\protect\alpha <0:$}\ 

In this case it is not possible to find a solution.

\subsection{Case $T_{\protect\mu \protect\nu }=0$ and $k=1 $}

Introducing (\ref{eqz08}) into (\ref{eqz06}) we obtain
\begin{equation}
6\left( \frac{{\dot{a}}^{2}+k}{a^{2}}\right) =3l^{2}\alpha \left( \frac{{%
\dot{a}}^{2}+k}{a^{2}}\right) ^{2}
\end{equation}%
which can be rewritten as

\begin{equation}  \label{eqz19}
\left( \frac{{\dot a}^{2}+k}{a^{2}}\right) \left( \frac {2}{\alpha l^{2}}- 
\frac{{\dot a}^{2}+k}{a^{2}}\right) =0.
\end{equation}

\subsubsection{Case $\ddot{a}=0$}

In this case it is not possible to find a solution.

\subsubsection{Case $\ddot{a}\neq 0$}

From equation (\ref{eqz19}) we obtain 
\begin{equation}
\frac{2}{\alpha l^{2}}a^{2}-{\dot{a}}^{2}=k.  \label{eqz20}
\end{equation}%
From (\ref{eqz20}) we can see two cases:

\paragraph{Case $\protect\alpha >0:$}\ 

\begin{figure}[h]
\includegraphics[width=350\unitlength]{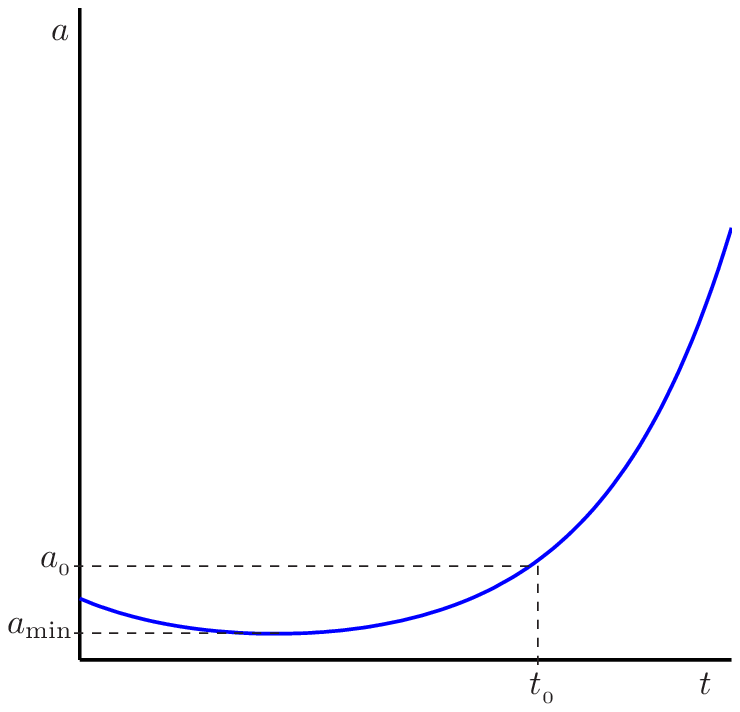}
\centering
\caption{ Graph of $a(t)$ with $\alpha>0$ and $k=1$. See equation (\ref{eqz21}).\label{fig05}}
\end{figure}

If $\alpha >0$ we can postulate a solution given by 
\begin{equation}
a(t)=A\cosh \left( \sqrt{\frac{2}{\alpha l^{2}}}(t-t^{\prime })\right)
\end{equation}%
where $t^{\prime }$ is a constant of integration, which leads 
\begin{equation}
A=\sqrt{\frac{\alpha l^{2}k}{2}}.
\end{equation}%
The initial condition $a_{0}=a(t=t_{0})$ leads%
\begin{equation}
a(t)=\sqrt{\frac{\alpha l^{2}k}{2}}\cosh \left[ \sqrt{\frac{2}{\alpha l^{2}}}%
(t-t^{\prime })+\textrm{arcosh} \left( \sqrt{\frac{2}{\alpha l^{2}}}a_{0}\right) %
\right]  \label{eqz21}
\end{equation}%
and%
\begin{equation}
\dot{a}(t)=\sqrt{k}\sinh \left[ \sqrt{\frac{2}{\alpha l^{2}}}%
(t-t_{0})+\textrm{arcosh}  \left( \sqrt{\frac{2}{\alpha l^{2}}}a_{0}\right) \right]
\label{eqz22}
\end{equation}%
which shows an accelerated expansion (see Fig. \ref{fig05})

Replacing (\ref{eqz21}) and (\ref{eqz22}) into (\ref{eqz06} - \ref{eqz09}) we
obtain%
\begin{equation*}
\rho ^{(h)}=-p^{(h)}=\frac{12}{\kappa _{5}\alpha l^{2}}
\end{equation*}%
i.e., we have an accelerated expansion when the energy density is positive
and pressure is negative (like a cosmological constant positive)

From equation (\ref{eqz10}) we find
\begin{equation}
-\frac{\dot{h}}{h-h(0)}=\frac{\dot{a}}{a},
\end{equation}%
so that 
\begin{equation}
h(t)=\frac{C}{\cosh \left[ \sqrt{\frac{2}{\alpha l^{2}}}(t-t_{0})+\text{%
arcosh}\left( \sqrt{\frac{2}{\alpha l^{2}k}}\,a_{0}\right) \right] }+h(0)
\end{equation}%
where $C$ is a constant of integration. The initial condition $%
h_{0}=h(t_{0}) $ leads
\begin{equation*}
h(t)=\frac{(h_{0}-h(0))\sqrt{\frac{2}{\alpha l^{2}k}}\,a_{0}}{\cosh \left[ 
\sqrt{\frac{2}{\alpha l^{2}}}(t-t_{0})+\text{arcosh}\left( \sqrt{\frac{2}{%
\alpha l^{2}k}}\,a_{0}\right) \right] }+h(0)
\end{equation*}%
from where we can see that $h(t)\rightarrow h(0)$ when $t\rightarrow \infty$.
\newline

\paragraph{Case $\protect\alpha <0:$}\ 

If $\alpha<0$ the equation (\ref{eqz20}) have no solution.

\subsection{Era of Dark Energy from Einstein-Chern-Simons gravity}

The results in the previous section are summarized in Tables \ref{table01}, \ref{table02} and \ref{table03}.

\begin{table}[h]
\centering
\caption{Solutions for scale factor of  an open space $k=-1$ (hyperbolic).}\label{table01}
\begin{tabular*}{\columnwidth}{@{\extracolsep{\fill}}lllll@{}}
\hline
Dynamics&$\alpha$&$\rho^{(h)}$&$p^{(h)}$&$\Lambda$\\
$a(t)$&&&&compatible\\
\hline
Accelerated&$>0$&$>0$&$<0$&$>0$\\
Decelerated&$<0$&$<0$&$>0$&$<0$\\
No accelerated&\multirow{2}{*}{any}&\multirow{2}{*}{$0$}&\multirow{2}{*}{$0$}&\multirow{2}{*}{$-$}\\
(\textit{Vacuum})\\
\hline
\end{tabular*}
\end{table}

\begin{table}[h]
\centering
\caption{Solutions for scale factor of  a flat space $k=0$.}\label{table02}
\begin{tabular*}{\columnwidth}{@{\extracolsep{\fill}}lllll@{}}
\hline
Dynamics&$\alpha$&$\rho^{(h)}$&$p^{(h)}$&$\Lambda$\\
$a(t)$&&&&compatible\\
\hline
Accelerated&$>0$&$>0$&$<0$&$>0$\\
Stationary&\multirow{2}{*}{any}&\multirow{2}{*}{$0$}&\multirow{2}{*}{$0$}&\multirow{2}{*}{$-$}\\
(\textit{Vacuum})\\
\hline
\end{tabular*}
\end{table}

\begin{table}[h]
\centering
\caption{Solutions for scale factor of  a closed space $k=1$.}\label{table03}
\begin{tabular*}{\columnwidth}{@{\extracolsep{\fill}}lllll@{}}
\hline
Dynamics&$\alpha$&$\rho^{(h)}$&$p^{(h)}$&$\Lambda$\\
$a(t)$&&&&compatible\\
\hline
Accelerated&$>0$&$>0$&$<0$&$>0$\\
\hline
\end{tabular*}
\end{table}

So that we have found solutions that describe accelerated expansion
for the three possible cosmological models of the universe. Namely,
spherical expansion $\left( k=1\right) $, flat expansion $\left(
k=0\right) $ and hyperbolic expansion $\left( k=-1\right) $ when the constant $\alpha $ is greater than zero. This means that
the Einstein-Chern-Simons field equations have as a of their solutions a
universe in accelerated expansion. This result allow us to conjeture that
this solutions are compatible with the era of Dark Energy and that the
energy-momentum tensor for the field $h^{a}$ corresponds to a form
of positive cosmological constant. 

From this solutions we can see that as time passes, the $h(t)$ decreases rapidly to $h(0)$, a constant value, keeping constant matter density.

We have also shown that the EChS field equations have
solutions that allows us to identify the energy-momentum tensor for the
field $h^{a}$ with a negative cosmological constant.

\section{Consistency of the Solutions with the "Era of Matter"\label{sec04}}

In the previous section, we find that the solutions of EChS field equations, with $T_{\mu\nu}=0$, can be useful as models of the era of Dark Energy. In this section we review the consistency of this equations with the era of Matter.

We will consider the ordinary matter as dust ($\rho\neq 0$, $p = 0$), such as occurs in standard cosmology. The non-ordinary matter will be
modeled as a perfect fluid ($\rho^{(h)}\neq 0$ y $p^{(h)}\neq 0$). In this case the field equations (\ref{eqz01} - \ref{eqz05}) takes the form

\begin{align}
6\left( \frac{{\dot{a}}^{2}+k}{a^{2}}\right)^{\phantom{B}} &=\kappa _{5}\left( \rho +\alpha\rho ^{(h)}\right)  \label{eqz23}\\
3\left[ \frac{\ddot{a}}{a}+\left( \frac{{\dot{a}}^{2}+k}{a^{2}}\right)\right]^{\phantom{B}} &=-\kappa _{5}\alpha p^{(h)}  \label{eqz24}\\
{\frac{3l^{2}}{\kappa _{5}}\left( \frac{{\dot{a}}^{2}+k}{a^{2}}\right)^{2}}^{\phantom{B}}&=\rho ^{(h)}  \label{eqz25}\\
\frac{3l^{2}}{\kappa _{5}}\frac{\ddot{a}}{a}\left( \frac{{\dot{a}}^{2}+k}{a^{2}}\right)^{\phantom{B}} &=-p^{(h)}  \label{eqz26}\\
\left( \frac{{\dot{a}}^{2}+k}{a^{2}}\right) \left[ (h-h(0))\frac{\dot{a}}{a}+\dot{h}\right]^{\phantom{B}} &=0  \label{eqz27}
\end{align}%
and the conservation equations (\ref{ect01}) (\textit{di\-ver\-gen\-ce-free ener\-gy-mo\-men\-tum ten\-sor}) for each fluids are given by 
\begin{equation}
\dot{\rho}+4\frac{\dot{a}}{a}\rho =0  \label{eqz28}
\end{equation}
and
\begin{equation}
\dot{\rho}^{(h)}+4\frac{\dot{a}}{a}\left( \rho ^{(h)}+p^{(h)}\right) =0.
\label{eqz29}
\end{equation}

The equation (\ref{eqz28}) have as solution  
\begin{equation}
\rho (t)=\left( \frac{a_{0}}{a(t)}\right) ^{4}\rho _{0}  \label{eqz30}
\end{equation}
where the initial conditions $a_{0}=a(t_{0}) $ and $\rho _{0}=\rho (t_{0})$ has been set.

Replacing (\ref{eqz30}) and (\ref{eqz25}) into equation (\ref{eqz23}) we have
\begin{equation}
\left( \frac{{\dot{a}}^{2}+k}{a^{2}}\right) ^{2}-2A\left( \frac{{\dot{a}}%
^{2}+k}{a^{2}}\right) +AB\,\frac{a_{0}^{4}}{a^{4}}=0  \label{eqz33}
\end{equation}%
where we defined
\begin{equation}
A:=\frac{1}{\alpha l^{2}}\quad ,\quad B:=\frac{\kappa _{5}\rho _{0}}{3}.
\label{eqz32}
\end{equation}

\subsection{Case $k=-1$\label{sub3.1}}

In this case, the equation (\ref{eqz33}) can be rewritten 
\begin{equation}
\left( \frac{{\dot{a}}^{2}-1}{a^{2}}\right) ^{2}-2A\left( \frac{{\dot{a}}%
^{2}-1}{a^{2}}\right) +AB\,\frac{a_{0}^{4}}{a^{4}}=0  \label{eqz34}
\end{equation}%
where we find
\begin{equation}
\dot{a}=\pm \sqrt{Aa^{2}\left( 1\pm \text{sgn}(A)\sqrt{1-\frac{B}{A}\,\frac{%
a_{0}^{4}}{a^{4}}}\right) +1} . \label{eqz35}
\end{equation}

\subsubsection{Case $\protect\alpha >0$}

In this case 
\begin{equation}
A=\frac{1}{\alpha l^{2}}>0.
\end{equation}
From (\ref{eqz35}) we can see that $\dot{a}$ is well defined if 
\begin{equation}
a\geq a_{\text{min}}=\sqrt[4]{\frac{B}{A}}\,a_{0}  \label{eqz36}
\end{equation}%
where%
\begin{equation*}
a_{\text{min}}=\sqrt[4]{\frac{\kappa _{5}\alpha l^{2}\rho _{0}}{3}}\,a_{0}.
\end{equation*}

On the other hand $a_0$ must satisfy
\begin{equation}
a_{0}\geq a_{\text{min}}
\end{equation}%
so that 
\begin{equation}
\frac{B}{A}\leq 1\quad \text{i.e.,}\quad B\leq A  \label{eqz37}
\end{equation}%
and therefore%
\begin{equation}
\rho _{0}\leq \rho _{\max }=\frac{3}{\kappa _{5}\alpha l^{2}}
\end{equation}

These results allow us to analyze the radicand in (\ref{eqz35}) 
\begin{equation}
Aa^{2}\left( 1\pm \sqrt{1-\frac{a_{\text{min}}^{4}}{a^{4}}}\right) +1\geq
0, 
\end{equation}
i.e.,
\begin{equation}
-A^{2}a_{\text{min}}^{4}\leq 1+2Aa^{2}
\end{equation}
which is satisfied for all $a$.
\newline

\paragraph{Plus or minus sign?}\ 

The choice of the sign into the radicand has information about the allowed
values of $\dot{a}$. Let us consider $\dot{a}>0$ (the analysis of the case $\dot{a}<0$ is very
similar) 
\begin{equation}
\dot{a}=\sqrt{Aa^{2}\left( 1\pm \sqrt{1-\frac{a_{\text{min}}^{4}}{a^{4}}}%
\right) -k} . \label{143}
\end{equation}
The function $\dot{a}(a)$ is monotonically increasing (decreasing) if we
consider the plus (minus) sign in front of the square root.\newline

From (\ref{143}) we can see that there exist $\dot{a}_{\text{cri}}$%
\begin{equation}
\dot{a}_{\text{cri}}:=\dot{a}(a_{\text{min}}^{4})=\sqrt{\sqrt{\frac{\kappa _{5}\rho _{0}}{3\alpha l^{2}}}a_{\text{0}}^{2}-k}.
\end{equation}

If we consider the plus (minus) sign in front of the square root, $\dot{a}_{%
\text{cri}}$ is the minimum (maximum) value of $\dot{a}$.

\begin{figure}[h]
\includegraphics[width=350\unitlength]{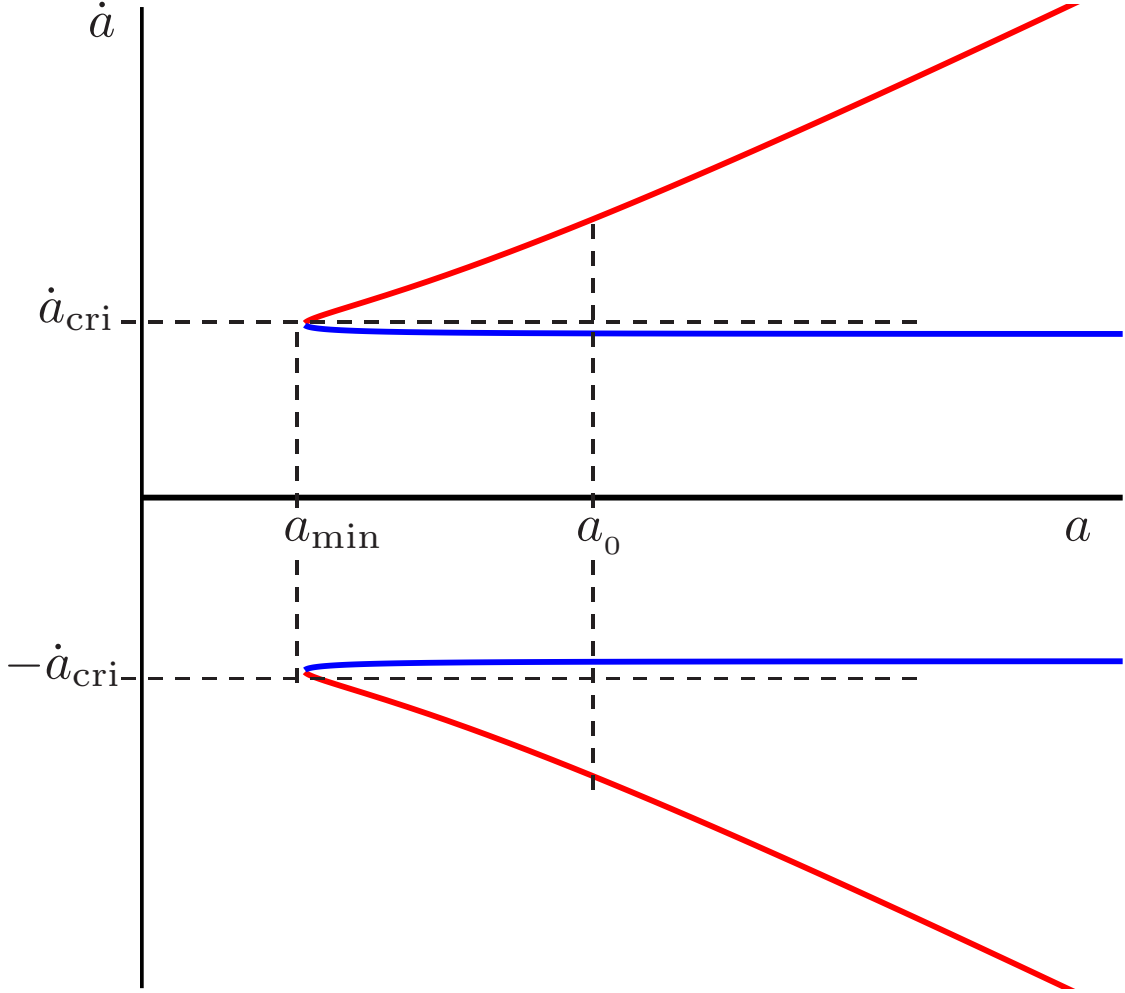}
\centering
\caption{ For every $a_0$ allowed there are two different values for $\dot a>0$: evolution with $\dot a$ approximate constant and evolu\-tion acce\-le\-ra\-ted(decelerated).\label{fig06}}
\end{figure}

If there is a limit to $a\gg a_{\text{min}},$ then

\begin{align} 
\dot a&=\pm\sqrt{Aa^2\left(1\pm\sqrt{1-\frac{a_\text{min}^4}{a^4}} \right)-k}\notag\\
&\approx\pm\sqrt{Aa^2\left\{1\pm\left(1-\frac{a_\text{min}^4}{2a^4}\right)\right\}-k} \label{eqz38}
\end{align}
where $k=-1$.
\newline

\paragraph{Case where the sign is ``$+$''}\ 

In this case%
\begin{equation}
\dot{a}=\pm \sqrt{Aa^{2}\left( 2-\frac{a_{\text{min}}^{4}}{2a^{4}}\right) -k}%
\approx \pm \sqrt{2Aa^{2}-k},
\end{equation}%
whose approximate solution is%
\begin{align}
a(t)&=\pm \sqrt{-\frac{\alpha l^{2}k}{2}}\notag\\
&\quad \times\sinh \Biggl[ \sqrt{\frac{2}{\alpha
l^{2}}}(t-t_{0})+\textrm{arsinh}\left( \sqrt{-\frac{2}{\alpha l^{2}k}}\,a_{0}\right) %
\Biggr] 
\end{align}%
where we use $A=\frac{1}{\alpha l^{2}}$ and $k=-1$.
\newline

\begin{figure}[h]
\includegraphics[width=350\unitlength]{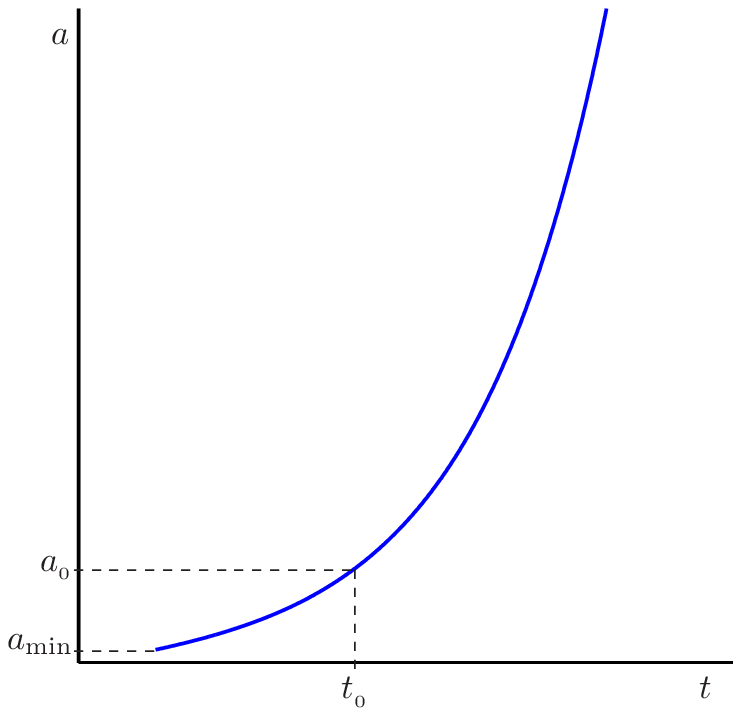}
\centering
\caption{Numerical solution with $A>0$, $k=-1$ and $\dot a_{_0}>\dot a_\text{cri}$ 
of \mbox{$\dot a=\sqrt{Aa^2\left(1+\sqrt{1-\frac{a_\text{min}^4}{a^4}} \right)-k}$} .\label{fig07}}
\end{figure}

\paragraph{Case where the sign is ``$-$''}\ 

In this case%
\begin{equation}
\dot{a}=\pm \sqrt{A\frac{a_{\text{min}}^{4}}{2a^{2}}-k}\approx \pm \sqrt{-k}
\end{equation}%
whose approximate solution is%
\begin{equation*}
a(t)=\pm \sqrt{-k}(t-t_{0})+a_{0}
\end{equation*}%
where we use $k=-1$.

\begin{figure}[h]
\includegraphics[width=350\unitlength]{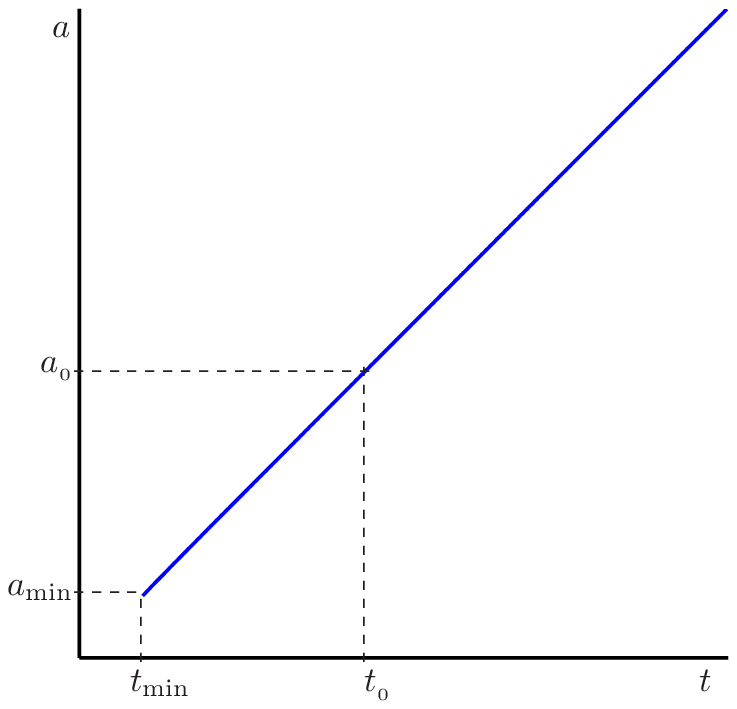}
\centering
\caption{Numerical solution with $A>0$, $k=-1$ and $\dot a_{_0}<\dot a_\text{cri}$ of \mbox{$\dot a=\sqrt{Aa^2\left(1+\sqrt{1-\frac{a_\text{min}^4}{a^4}} \right)-k}$} .\label{fig08}}
\end{figure}

\subsubsection{Case $\protect\alpha <0$}

In this case 
\begin{equation}
A=\frac{1}{\alpha l^2}<0.
\end{equation}

From (\ref{eqz35}) we can see that $\dot{a}$ is well defined if 
\begin{equation}
1-\frac{B}{A}\,\frac{a_{0}^{4}}{a^{4}}\geq 0,
\end{equation}%
but this condition is satisfied for all $a$.\newline

\paragraph{Case where the sign is `$+$'}\ 

In this case%
\begin{equation}
Aa^{2}\left( 1+\sqrt{1-\frac{Ba_{0}^{4}}{Aa^{4}}}\right) -k\geq 0,
\end{equation}%
so that 
\begin{equation}
\frac{k-Aa^{2}}{Aa^{2}}\geq \sqrt{1-\frac{Ba_{0}^{4}}{Aa^{4}}}.
\label{eqz38.1}
\end{equation}
The left side of the last equation must be positive, i.e., 
\begin{equation}
k-Aa^{2}\leq 0\quad or\quad a\leq \sqrt{\frac{k}{A}}  .\label{eqz38.2}
\end{equation}%
From (\ref{eqz38.1}) we obtain 
\begin{equation}
k^{2}-2Aka^{2}\geq -ABa_{0}^{4}  \label{eqz38.3}
\end{equation}%
and again, the left side of the last equation must be positive, i.e., 
\begin{equation}
k^{2}-2Aka^{2}\geq 0\quad \Longleftrightarrow \quad a\leq \sqrt{\frac{k}{2A}}
\end{equation}%
and from (\ref{eqz38.3}) we find 
\begin{equation}
a\leq \sqrt{\frac{k^{2}+ABa_{0}^{4}}{2Ak}}.  \label{eqz38.4}
\end{equation}%
Since 
\begin{equation}
\sqrt{\frac{k}{A}}>\sqrt{\frac{k}{2A}}>\sqrt{\frac{k^{2}+ABa_{0}^{4}}{2Ak}}%
=a_{\text{max}}\geq a,
\end{equation}%
we have found a maximum value for $a$%
\begin{equation}
a_{\max }=\sqrt{\frac{3\alpha l^{2}k^{2}+\kappa _{5}\rho _{0}a_{0}^{4}}{6k}}
\end{equation}%
and therefore 
\begin{equation}
\dot{a}(a=a_{\text{max}})=0
\end{equation}%
i. e., $a_{\text{max}}$ is a local maximum. It is direct to prove that $\dot{%
a}\neq 0$ for $a\neq a_{\text{max}}$.
If $a$ has a maximum value $a_{\text{max}}$ then (see (\ref{eqz30})) 
\begin{equation}
\rho (t)=\left( \frac{a_{0}}{a(t)}\right) ^{4}\rho _{0}\geq \left( \frac{%
a_{0}}{a_{\text{max}}}\right) ^{4}\rho _{0}=\rho _{\text{min}}.
\end{equation}%
This means that $\rho $ has a minimum value $\rho _{\text{min}}$ given by%
\begin{equation}
\rho _{\text{min}}=\left( \frac{6ka_{0}^{2}}{3\alpha l^{2}k^{2}+\kappa
_{5}\rho _{0}a_{0}^{4}}\right) ^{2}\rho _{0}
\end{equation}%
where $k=-1$.

\begin{figure}[h]
\includegraphics[width=350\unitlength]{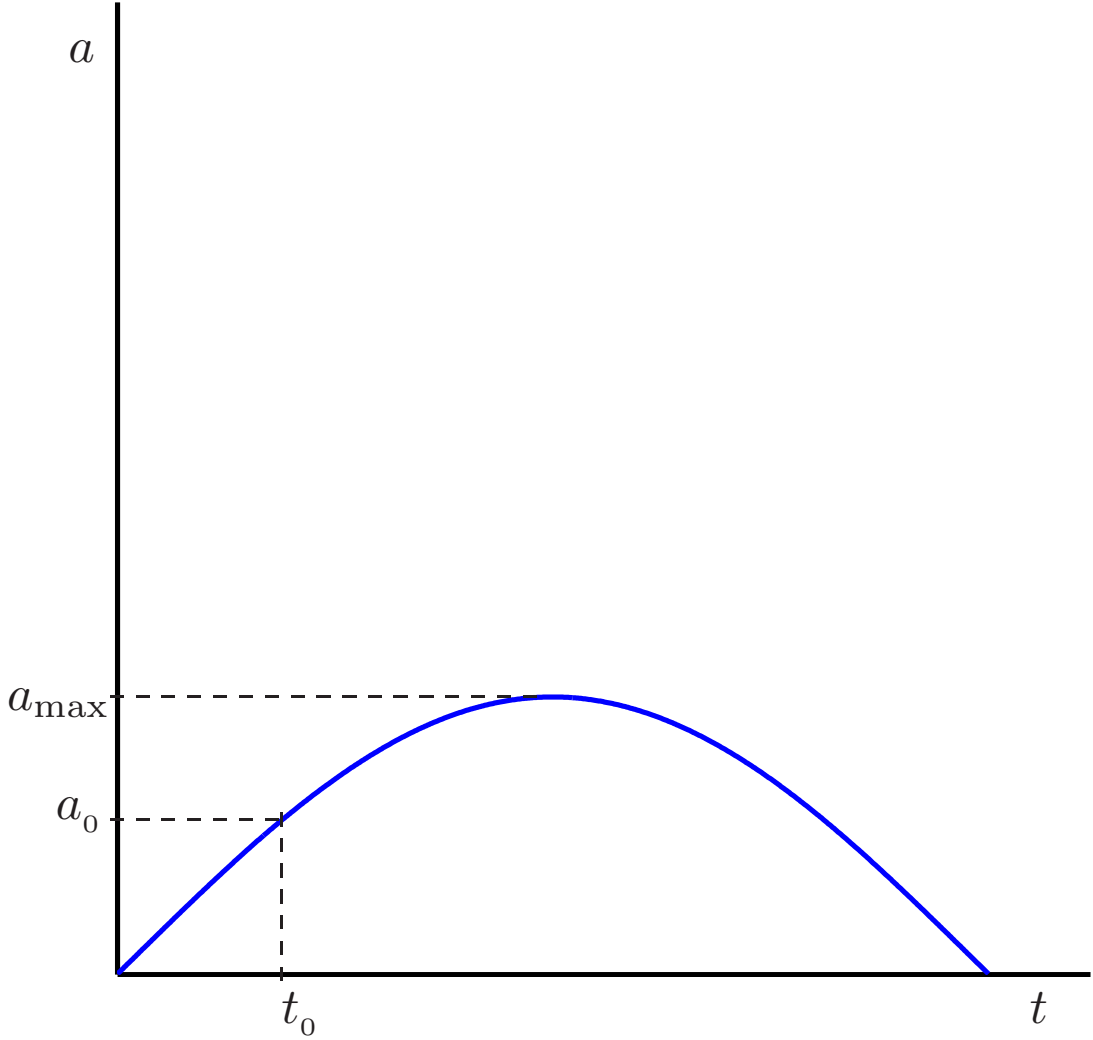}
\centering
\caption{ Solution of $\dot a=\sqrt{Aa^2\left(1-\sqrt{1-\frac{Ba_0^4}{Aa^4}} \right)-k}$ with $A<0$, $k=-1$ and $|\dot a_{_0}|<\dot a_\text{max}$.\label{fig09}}
\end{figure}

Consider the case where $\dot{a}>0.$ We just consider $\dot{a}>0$ because the analysis of the case $\dot{a}<0$
looks very similar.  In this case

\begin{equation}
\dot{a}=\sqrt{Aa^{2}\left( 1+\sqrt{1-\frac{Ba_{0}^{4}}{Aa^{4}}}\right) -k}
\end{equation}%
is a decreasing function. We can see that the minimum value of $\dot{a}$ is
given by%
\begin{equation}
\dot{a}_{\min }=\dot{a}(a_{\max })=0
\end{equation}%
and the maximum value of $\dot{a}$ is given by%
\begin{equation}
\dot{a}_{\max }=\dot{a}(a=0)=\sqrt{-\sqrt{-\frac{\kappa _{5}\rho _{0}}{3\alpha l^{2}}}a_{0}^{2}-k}.\notag
\end{equation}

\begin{figure}[h]
\includegraphics[width=350\unitlength]{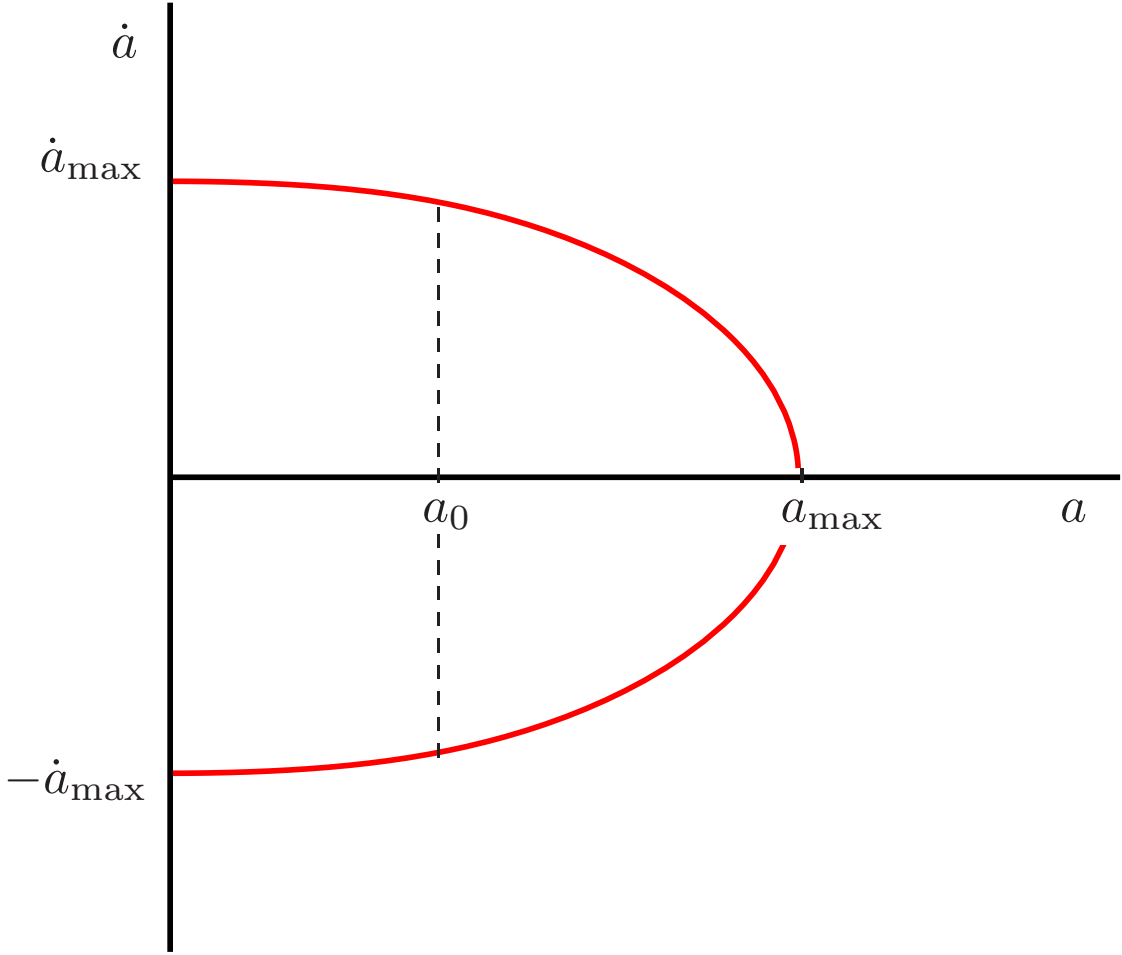}
\centering
\caption{Phase space for $A<0$ and $k=-1$ with ``$+$'' sign.\label{fig10}}
\end{figure}

\paragraph{Case where the sign is ``$-$''}\ 

In this case we obtain the following condition

\begin{equation}
Aa^2\left(1-\sqrt{1-\frac{Ba_0^4}{Aa^4}} \right)-k\geq 0
\end{equation}
where $A=-\frac{1}{\alpha l^2}$ and $k=-1$. This condition is trivially
satisfied for all $a$.\newline

This result implies that $\dot{a}\neq 0$. This means that $a$ has no local
maximums/minimums, so $a$ is monotonically increasing or monotonically
decreasing.

If there is a limit to $a\gg \sqrt[4]{-\frac{B}{A}}\,a_{0},$ then
\begin{align}
\dot{a}&=\pm \sqrt{Aa^{2}\left( 1-\sqrt{1-\frac{Ba_{0}^{4}}{Aa^{4}}}\right) -k}\notag\\
&\approx \pm \sqrt{Aa^{2}\left( 1-\left( 1-\frac{Ba_{0}^{4}}{2Aa^{4}}\right)\right) -k}
\end{align}%
and 
\begin{equation}
\dot{a}=\pm \sqrt{\frac{Ba_{0}^{4}}{2a^{2}}-k}\approx \pm \sqrt{-k},
\end{equation}%
whose approximate solution is%
\begin{equation}
a(t)=\pm \sqrt{-k}(t-t_{0})+a_{0}
\end{equation}%
where we use $k=-1$.

\begin{figure}[h]
\includegraphics[width=350\unitlength]{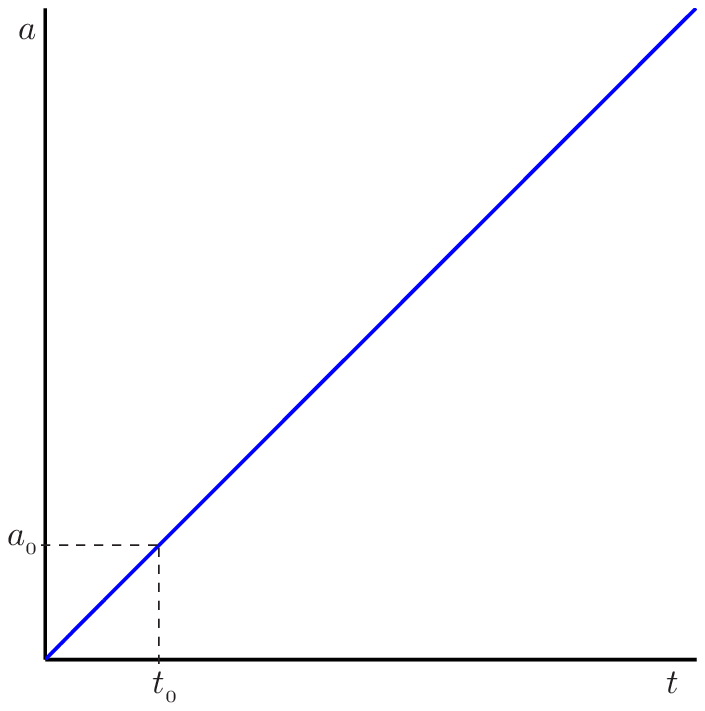}
\centering
\caption{Solution of $\dot a=\sqrt{Aa^2\left(1-\sqrt{1-\frac{Ba_0^4}{Aa^4}} \right)-k}$ with $A<0$, \mbox{$k=-1$} and $1<\dot a_{_0}<\dot a_\text{max}$.\label{fig11}}
\end{figure}

In this case
\begin{equation}
\dot{a}=\sqrt{Aa^{2}\left( 1-\sqrt{1-\frac{Ba_{0}^{4}}{Aa^{4}}}\right) -k}
\end{equation}%
is a decreasing function. The maximum value of $\dot{a}$ is given by%
\begin{equation}
\dot{a}_{\max }=\dot{a}(a=0)=\sqrt{\sqrt{-\frac{%
\kappa _{5}\rho _{0}}{3\alpha l^{2}}}a_{0}^{2}-k}
\end{equation}%
and we can see that $\dot{a}$ tends to a minimum value given by%
\begin{equation}
\dot{a}_{\min }=\dot{a}(a\rightarrow \infty )=\sqrt{-k}=1
\end{equation}

\begin{figure}[h]
\includegraphics[width=350\unitlength]{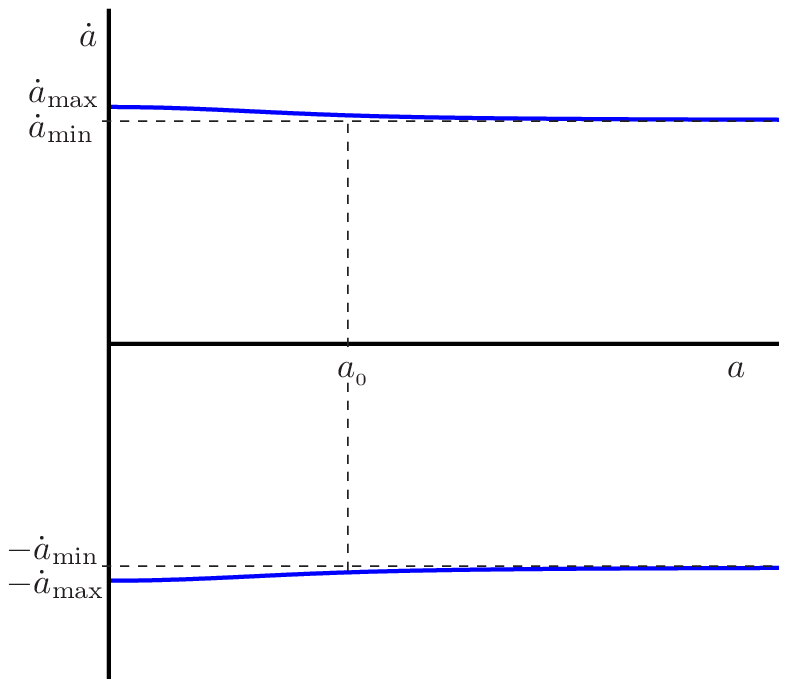}
\centering
\caption{Phase space for $A<0$ and $k=-1$ with ``$-$'' sign.\label{fig12}}
\end{figure}

\begin{figure}[h]
\includegraphics[width=350\unitlength]{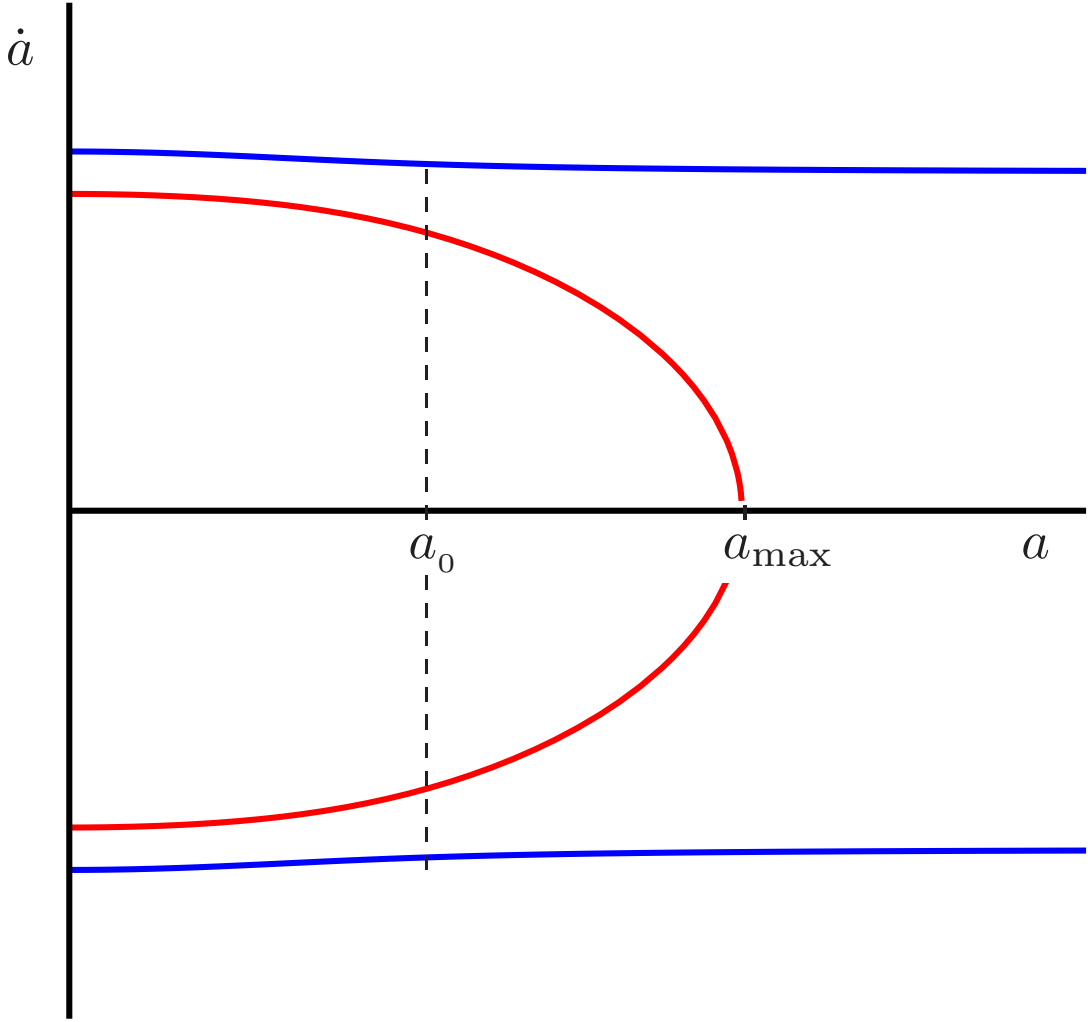}
\centering
\caption{Phase space for $A<0$ and $k=-1$. Comparison between phase space with ``$+$'' sign (Fig. \ref{fig10}) and ``$-$'' sign (Fig. \ref{fig12}).\label{fig13}}
\end{figure}

\subsection{Case $k=0$}

In this case, the equation (\ref{eqz33}) takes the form 
\begin{equation}
\left( \frac{\dot{a}}{a}\right) ^{4}-2A\left( \frac{\dot{a}}{a}\right)
^{2}+AB\,\frac{a_{0}^{4}}{a^{4}}=0
\end{equation}%
from where 
\begin{equation}
\dot{a}=\pm \sqrt{Aa^{2}\left( 1\pm \text{sgn}(A)\sqrt{1-\frac{B}{A}\,\frac{%
a_{0}^{4}}{a^{4}}}\right) }  .\label{eqz39}
\end{equation}

\subsubsection{Case $\protect\alpha >0$}

In this case 
\begin{equation}
A=\frac{1}{\alpha l^{2}}>0.
\end{equation}

From (\ref{eqz39}) we can see that $\dot{a}$ is well defined if 
\begin{equation}
a\geq \sqrt[4]{\frac{B}{A}}\,a_{0}
\end{equation}%
and therefore a minimum value for $a$ is given by%
\begin{equation}
a_{\min }=\sqrt[4]{\frac{\kappa _{5}\alpha l^{2}\rho _{0}}{3}}a_{0}.
\end{equation}

On the other hand $a_{0}\geq a_{\text{min}}$, so that 
\begin{equation}
B\leq A\quad \text{i.e.,}\quad \rho _{0}\leq \rho _{\max }=\frac{3}{\kappa
_{5}\alpha l^{2}}.
\end{equation}%
These results leads 
\begin{equation}
Aa^{2}\left( 1\pm \text{sgn}(A)\sqrt{1-\frac{B}{A}\,\frac{a_{0}^{4}}{a^{4}}}%
\right) \geq 0,
\end{equation}%
i.e., $a$ has no local maximums/minimums \footnote{Only it has a local maximum/minimum if we consider the minus sign into the radicand. In that case the local minimum is $a_{\text{min}}$. We can prove
that there is no local maximum.}, 
so that $a$ is mo\-no\-to\-ni\-ca\-lly increasing or
monotonically decreasing.
\newline

\paragraph*{Plus or minus sign?}\  

The choice of the sign into the radicand has information about the
allowed values of $\dot{a}$. Let us consider $\dot{a}>0$, the analysis of
the case $\dot{a}<0$ is very similar 
\begin{equation}
\dot{a}=\sqrt{Aa^{2}\left( 1\pm \sqrt{1-\frac{a_{\text{min}}^{4}}{a^{4}}}%
\right) }  .\label{179}
\end{equation}

The function $\dot{a}(a)$ is monotonically increasing(de\-crea\-sing) if we
consider the plus(minus) sign in front of the square root.\newline

From (\ref{179}) we can see that exist $\dot{a}_{\text{cri}}$%
\begin{equation*}
\dot{a}_{\text{cri}}:=\dot{a}_{\min }=\sqrt{A}\,a_{\min }=\sqrt[4]{\frac{%
\kappa _{5}\rho _{0}}{3\alpha l^{2}}}\,a_{0}.
\end{equation*}

If we consider the plus (minus) sign in front of the square root, $\dot{a}_{%
\text{cri}}$ is the minimum(maximum) value of $\dot{a}$.
\newline

\begin{figure}[h]
\includegraphics[width=350\unitlength]{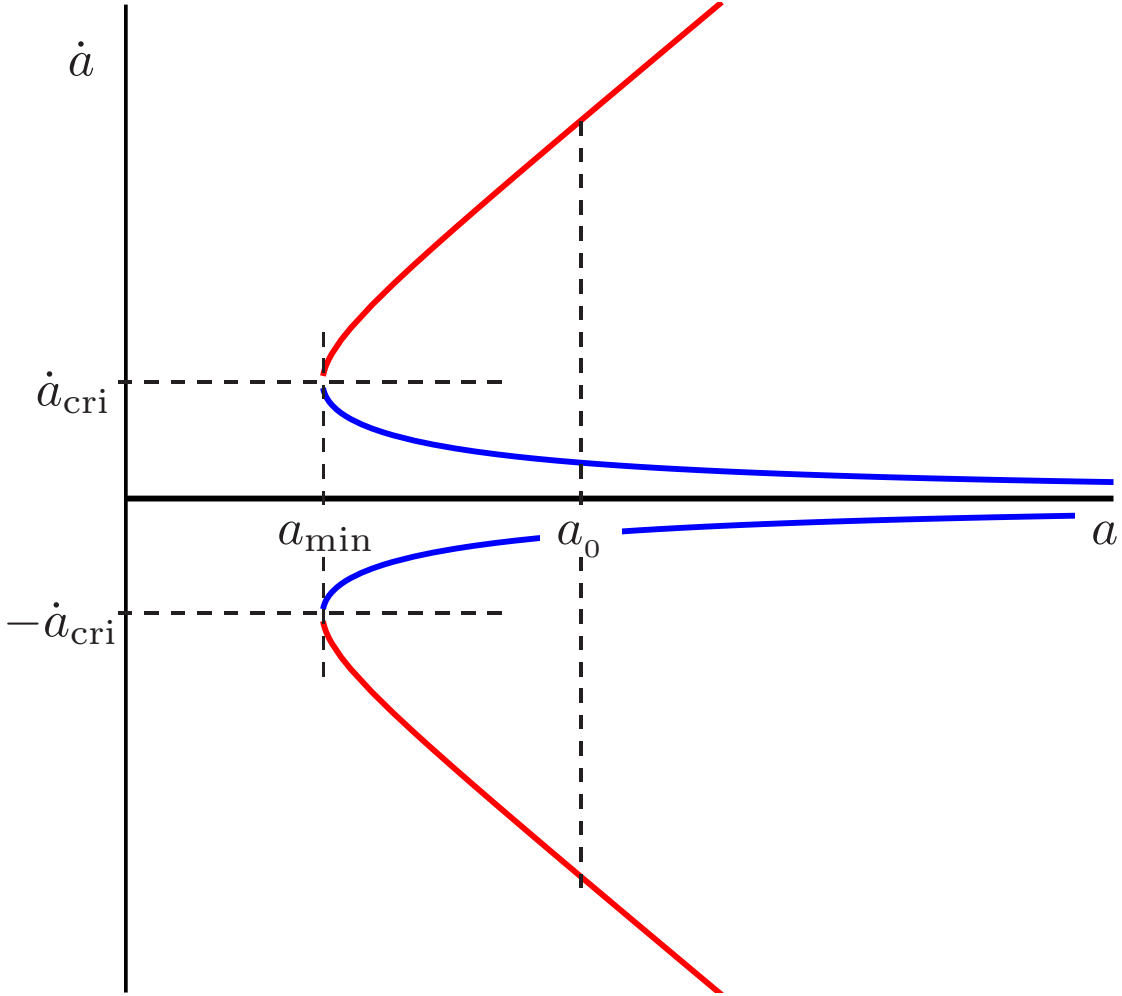}
\centering
\caption{For every $a_0$ there are two different values for $\dot a$: evolution with $\dot a<\dot a_\text{cri}$ and expansion accelerated(decelerated) with $|\dot a|>|\dot a_\text{cri}|$ .\label{fig14}}
\end{figure}

If there is a limit to $a\gg a_{\text{min}}$ then
\begin{align}
\dot a&=\pm\sqrt{Aa^2\left(1\pm\sqrt{1-\frac{a_\text{min}^4}{a^4}} \right)}\notag\\
&\approx\pm a\sqrt{A\left(1\pm\left(1-\frac{a_\text{min}^4}{2a^4}\right)
\right)}
\end{align}

\paragraph{Case where the sign is ``$+$''}\ 

In this case%
\begin{equation}
\dot{a}=\pm a\sqrt{A\left( 2-\frac{a_{\text{min}}^{4}}{2a^{4}}\right) }%
\approx \pm a\sqrt{2A}
\end{equation}%
whose approximate solution is%
\begin{equation}
a(t)=a_{0}\exp \left( \pm \sqrt{\frac{2}{\alpha l^{2}}}(t-t_{0})\right)
\end{equation}%
where $A=\frac{1}{\alpha l^{2}}>0$.
\newline

\begin{figure}[h]
\includegraphics[width=350\unitlength]{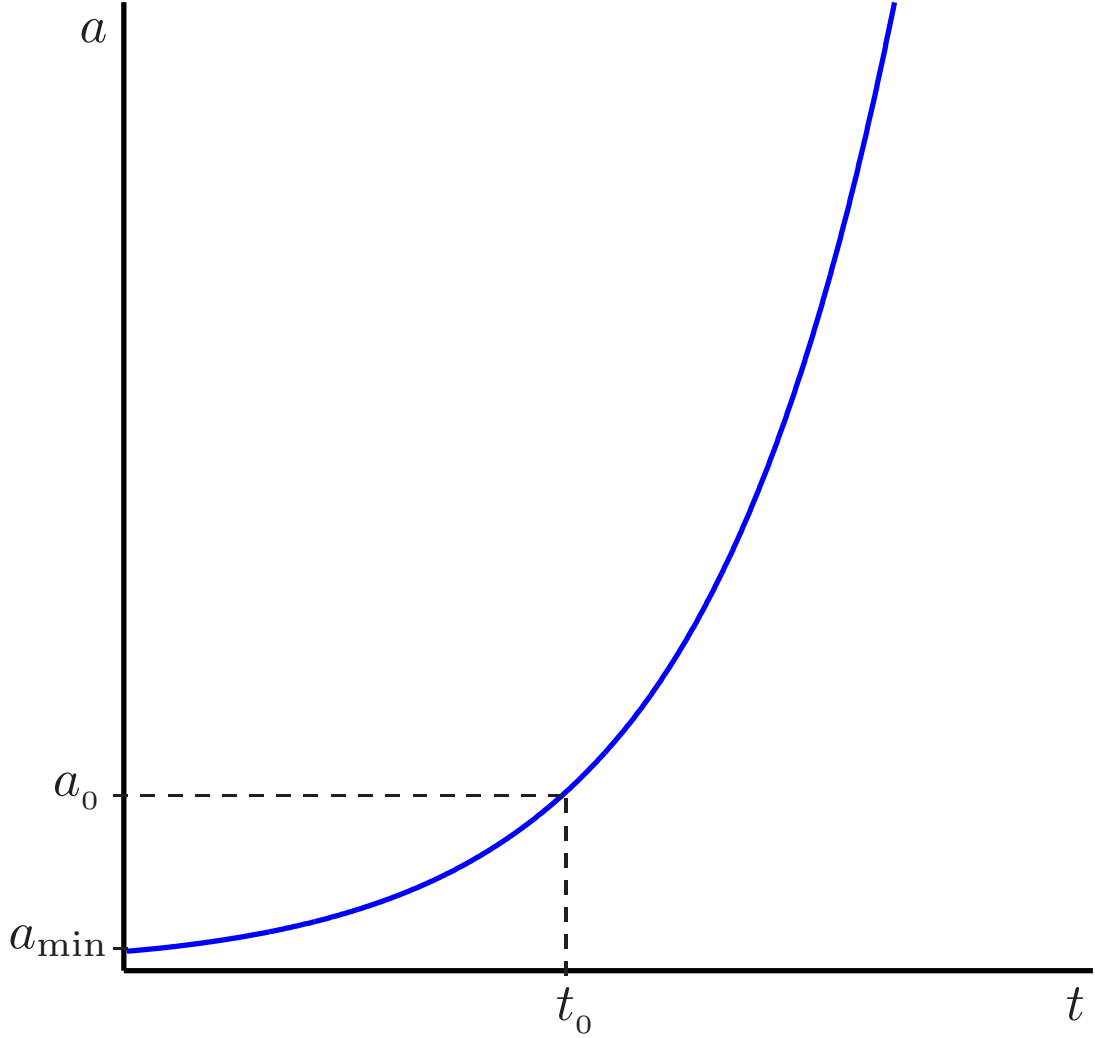}
\centering
\caption{Solution of $\dot a=\sqrt{Aa^2\left(1+\sqrt{1-\frac{a_\text{min}^4}{a^4}} \right)}$ with $A>0$ and \mbox{$\dot a_{_0}>\dot a_\text{cri}$}.\label{fig15}}
\end{figure}

\paragraph{Case where the sign is ``$-$''}\ 

In this case
\begin{equation}
\dot{a}\approx \pm \sqrt{\frac{A}{2}}\,\frac{a_{\text{min}}^{2}}{a}
\end{equation}%
whose approximate solution is%
\begin{align}
a(t)&=\pm \sqrt{a_{0}^{2}\pm \sqrt{\frac{2}{\alpha l^{2}}}a_{\text{min}%
}^{2}(t-t_{0})}\notag\\
&=\pm a_{0}\sqrt{1\pm \sqrt{\frac{2\kappa _{5}\rho _{0}}{%
3a_{0}^{4}}}(t-t_{0})}
\end{align}
where we use $A=\frac{1}{\alpha l^{2}}>0$ and $a_{\text{min}}=\sqrt[4]{\frac{%
\kappa _{5}\alpha l^{2}\rho _{0}}{3}}\,a_{0}$.


\begin{figure}[h]
\includegraphics[width=350\unitlength]{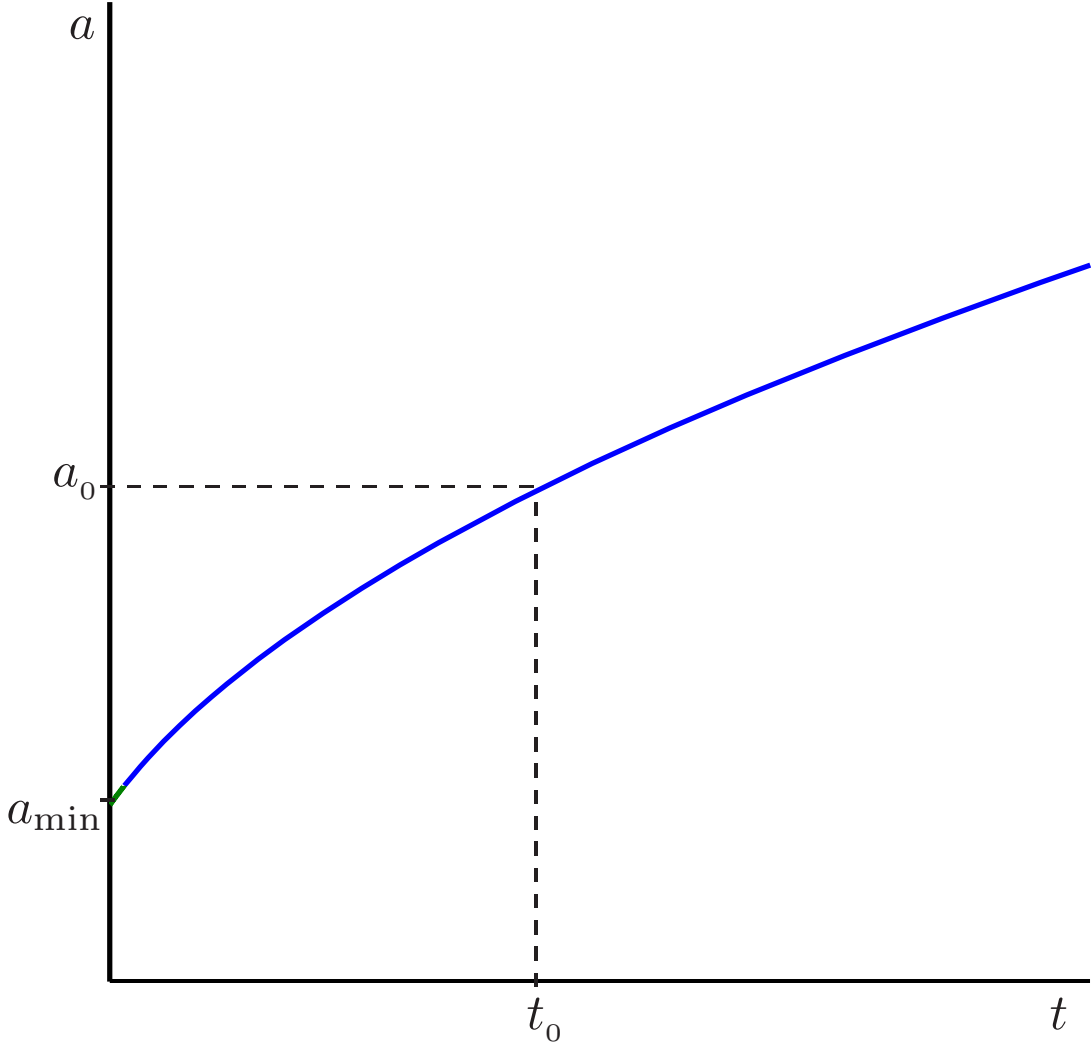}
\centering
\caption{ Solution of $\dot a=\sqrt{Aa^2\left(1-\sqrt{1-\frac{a_\text{min}^4}{a^4}} \right)}$ with $A>0$ and \mbox{$\dot a_{_0}<\dot a_\text{cri}$}.\label{fig16}}
\end{figure}

\subsubsection{Case $\protect\alpha <0$}

In this case 
\begin{equation}
A=\frac{1}{\alpha l^{2}}<0
\end{equation}%
From (\ref{eqz39}) we can see that $\dot{a}$ is well defined if 
\begin{equation}
1\mp \sqrt{1-\frac{B}{A}\,\frac{a_{0}^{4}}{a^{4}}}\leq 0.
\end{equation}%
This condition is only satisfied if we use the minus sign ``$-$'' for all $a$,
i.e., 
\begin{equation}
1-\sqrt{1-\frac{B}{A}\,\frac{a_{0}^{4}}{a^{4}}}<0
\end{equation}%
and therefore $a$ has no local maximums/minimums, so $a$ is monotonically
increasing or monotonically decreasing. \ So that $\dot{a}$ has a maximum
value in $a=0$, i.e., 
\begin{equation}
\dot{a}_{\max }=\dot{a}(a=0)=\sqrt[4]{-\frac{\kappa
_{5}\rho _{0}}{3\alpha l^{2}}}a_{0}
\end{equation}%
and $\dot{a}$ tends to a minimum value given by%
\begin{equation}
\dot{a}_{\min }=\dot{a}(a\rightarrow \infty )=0.
\end{equation}

\begin{figure}[h]
\includegraphics[width=350\unitlength]{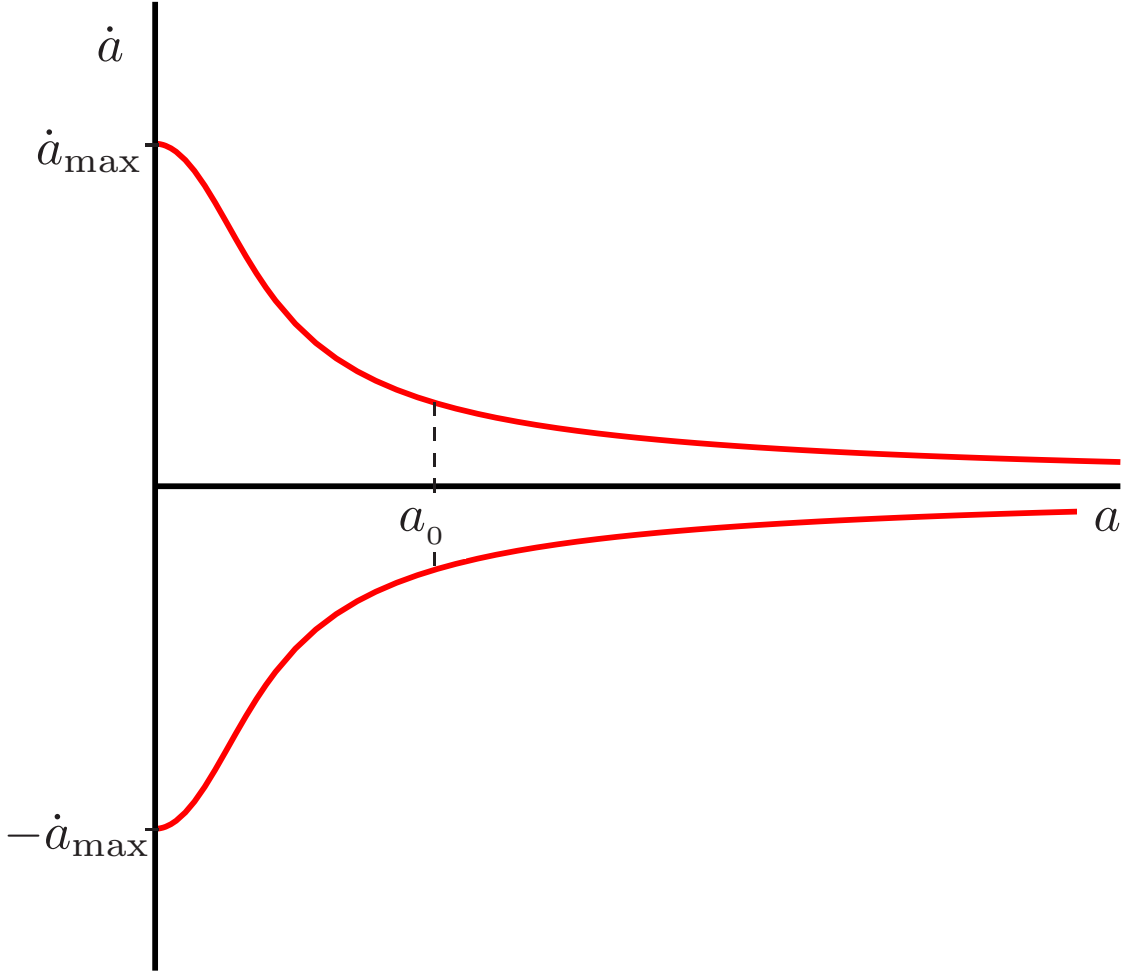}
\centering
\caption{Phase space for $A<0$ and $k=0$ with ``$-$'' sign.\label{fig17}}
\end{figure}

\begin{figure}[h]
\includegraphics[width=350\unitlength]{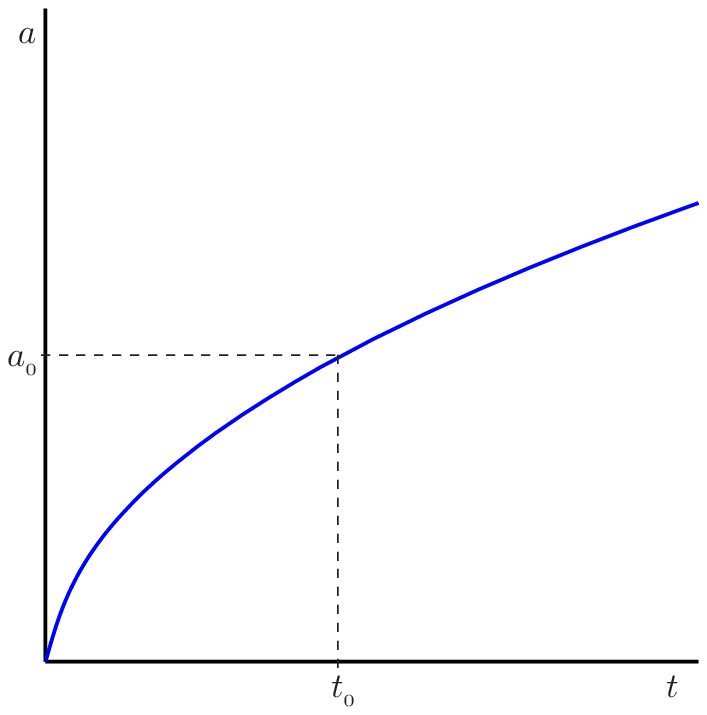}
\centering
\caption{Solution of $\dot a=\sqrt{Aa^2\left(1-\sqrt{1-\frac{Ba_0^4}{Aa^4}} \right)}$ with $A<0$, $k=0$ and $\dot a_{_0}<\dot a_\text{max}$.\label{fig18}}
\end{figure}

If exist a limit for $a\gg \sqrt[4]{-\frac{B}{A}}\,a_{0}$ then

\begin{equation}
\dot{a}=\pm \sqrt{Aa^{2}\left( 1-\sqrt{1-\frac{B}{A}\,\frac{a_{0}^{4}}{a^{4}}%
}\right) }\approx \pm \sqrt{\frac{B}{2}}\,\,\frac{a_{0}^{2}}{a},
\end{equation}%
whose approximate solution is%
\begin{equation}
a(t)=a_{0}\sqrt{1\pm \sqrt{\frac{2\kappa _{5}\rho _{0}}{3a_{0}^{4}}}(t-t_{0})}
\end{equation}%
where we use $B=\frac{\kappa _{5}\rho _{0}}{3}$.

\subsection{Case $k=1$}

In this case, the equation (\ref{eqz33}) can be rewritten as 
\begin{equation}
\left( \frac{{\dot{a}}^{2}+1}{a^{2}}\right) ^{2}-2A\left( \frac{{\dot{a}}%
^{2}+1}{a^{2}}\right) +AB\,\frac{a_{0}^{4}}{a^{4}}=0,  \label{eqz40}
\end{equation}%
from where 
\begin{equation}
\dot{a}=\pm \sqrt{Aa^{2}\left( 1\pm \text{sgn}(A)\sqrt{1-\frac{B}{A}\,\frac{%
a_{0}^{4}}{a^{4}}}\right) -k}  \label{eqz41}
\end{equation}%
with $k=1$.

\subsubsection{Case $\protect\alpha >0$}

In this case 
\begin{equation}
A=\frac{1}{\alpha l^{2}}>0.
\end{equation}%
From (\ref{eqz41}) we can see that $\dot{a}$ is well defined if%
\begin{equation}
a_{\min }=\sqrt[4]{\frac{\kappa _{5}\alpha l^{2}\rho _{0}}{3}}a_{0},
\end{equation}%
so that 
\begin{equation}
B\leq A\quad \text{i.e.,}\quad \rho _{0}\leq \rho _{\max }=\frac{3}{\kappa
_{5}\alpha l^{2}}.
\end{equation}

With these considerations we can analyze if the radicand is positive in (\ref{eqz41}) 
\begin{equation}  \label{eqz43}
Aa^2\left(1\pm\sqrt{1-\frac{a_\text{min}^4}{a^4}} \right)-k.
\end{equation}
\newline

\paragraph{Plus or minus sign?}\ 

Let us consider $\dot{a}>0$, the analysis of the case $\dot{a}<0$ is very similar 
\begin{equation}
\dot{a}=\sqrt{Aa^{2}\left( 1\pm \sqrt{1-\frac{a_{\text{min}}^{4}}{a^{4}}}%
\right) -k} . \label{200}
\end{equation}

The function $\dot{a}(a)$ is monotonically increasing (decreasing) if we
consider the plus (minus) sign in front of the square root.\newline

From (\ref{200}) we can see that exist $\dot{a}_{\text{cri}}$%
\begin{equation}
\dot{a}_\text{cri}:=\dot{a}(a_\text{min})=\sqrt{\sqrt{\frac{\kappa _{5}\rho _{0}}{3\alpha l^{2}}}a_{0}^{2}-k}.
\end{equation}%
If we consider the plus(minus) sign in front of the square root, $\dot{a}_{%
\text{cri}}$ is the minimum(maximum) value of $\dot{a}$.
\newline

\begin{figure}[h]
\includegraphics[width=350\unitlength]{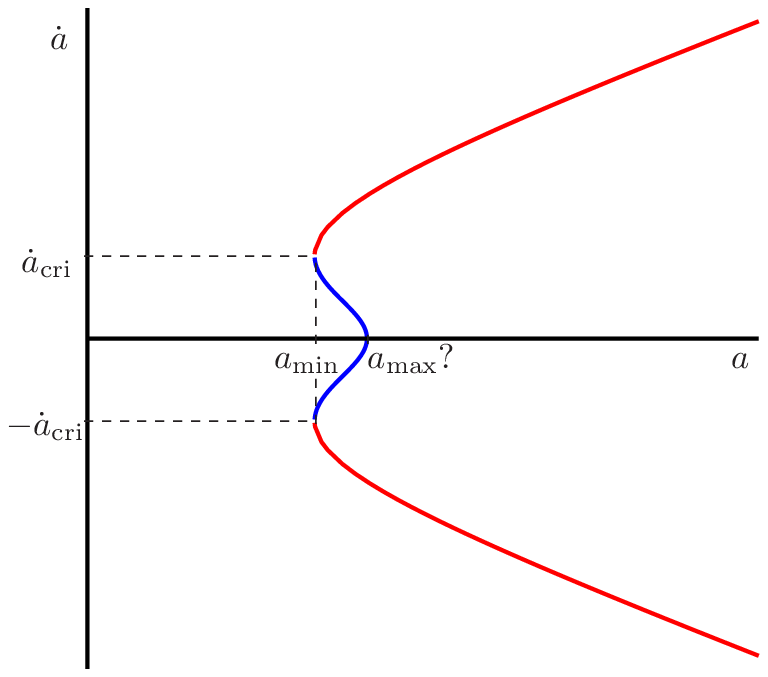}
\centering
\caption{Phase space for $A>0$ and $k=1$.\label{fig19}}
\end{figure}

\paragraph{Case where the sign is ``$+$''}\  

In this case
\begin{equation}
Aa^{2}\left( 1+\sqrt{1-\frac{a_{\text{min}}^{4}}{a^{4}}}\right) -k\geq Aa_{%
\text{min}}^{2}-k\geq 0,
\end{equation}%
so that 
\begin{equation}
a_{\text{min}}\geq \sqrt{\frac{k}{A}}\quad \Longleftrightarrow \quad \rho
_{0}a_{0}^{4}\geq 3\frac{\alpha l^{2}k^{2}}{\kappa _{5}},
\end{equation}%
but (see equation (\ref{eqz30})) 
\begin{equation*}
\rho (t)=\left( \frac{a_{0}}{a(t)}\right) ^{4}\rho _{0}\quad \Longrightarrow
\quad \rho a^{4}=\rho _{0}a_{0}^{4},
\end{equation*}%
then
\begin{equation}
\rho a^{4}\geq 3\frac{\alpha l^{2}k^{2}}{\kappa _{5}}.
\end{equation}

It is direct to prove that $\dot{a}\neq 0$ for $a>a_{\text{min}}$, then $a$
has no local maximums/minimums, and therefore $a$ is monotonically
increasing or monotonically decreasing.

If there is a limit to $a\gg a_{\text{min}}$ , then

\begin{equation}
\dot{a}=\pm \sqrt{Aa^{2}\left( 1+\sqrt{1-\frac{a_{\text{min}}^{4}}{a^{4}}}%
\right) -k}\approx \pm \sqrt{2Aa^{2}-k}
\end{equation}%
whose approximate solution is%
\begin{align*}
a(t)&=\pm \sqrt{\frac{\alpha l^{2}k}{2}}\\
&\quad\times\cosh \left[ \sqrt{\frac{2}{\alpha
l^{2}}}(t-t_{0})+\textrm{arcosh}\left( \sqrt{\frac{2}{\alpha l^{2}k}}a_{0}\right) %
\right]
\end{align*}%
where we use $A=\frac{1}{\alpha l^{2}}$ and $k=1$.

\begin{figure}[h]
\includegraphics[width=350\unitlength]{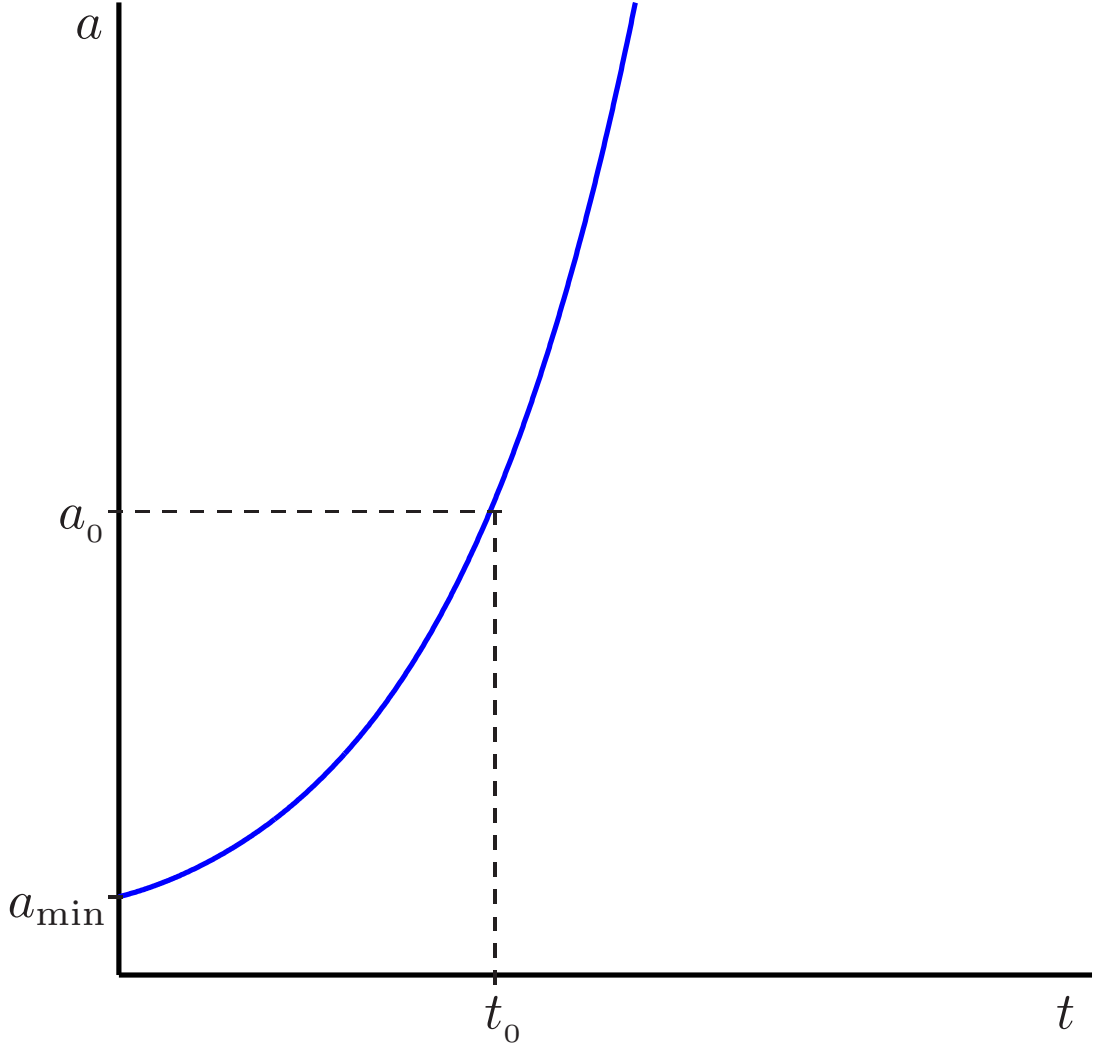}
\centering
\caption{Solution of $\dot a=\sqrt{Aa^2\left(1-\sqrt{1+\frac{a_\text{min}^4}{a^4}} \right)-k}$ with $A>0$, $k=1$ and $\dot a_{_0}>\dot a_\text{cri}$.\label{fig20}}
\end{figure}

\paragraph{Case where the sign is ``$\mathbf{-}$''}\ 

In this case%
\begin{equation}
Aa^{2}\left( 1-\sqrt{1-\frac{a_{\text{min}}^{4}}{a^{4}}}\right) -k\geq
0,
\end{equation}
therefore
\begin{equation}
\frac{Aa^{2}-k}{Aa^{2}}\geq \sqrt{1-\frac{%
a_{\text{min}}^{4}}{a^{4}}} . \label{eqz42}
\end{equation}
This condition must be also satisfied by $a_{\text{min}}$ 
\begin{equation}
Aa_{\text{min}}^{2}-k\geq 0\quad \Longleftrightarrow \quad a_{\text{min}%
}\geq \sqrt{\frac{k}{A}},
\end{equation}%
so that,
\begin{equation}
\rho _{0}a_{0}^{4}\geq 3\frac{\alpha l^{2}k^{2}}{\kappa _{5}},
\end{equation}%
but (see equation (\ref{eqz30})) 
\begin{equation*}
\rho a^{4}=\rho _{0}a_{0}^{4}
\end{equation*}%
and therefore%
\begin{equation}
\rho a^{4}\geq 3\frac{\alpha l^{2}k^{2}}{\kappa _{5}}.
\end{equation}

From (\ref{eqz42}) we obtain 
\begin{equation}
a\leq a_{\text{max}}=\sqrt{\frac{k^{2}+A^{2}a_{\text{min}}^{4}}{2Ak}},
\end{equation}
i.e.,
\begin{equation}
a_{\text{max}}=\sqrt{\frac{3\alpha l^{2}k^{2}+\kappa _{5}\rho
_{0}a_{0}^{4}}{6k}}.  \label{210}
\end{equation}%

\begin{figure}[h]
\includegraphics[width=350\unitlength]{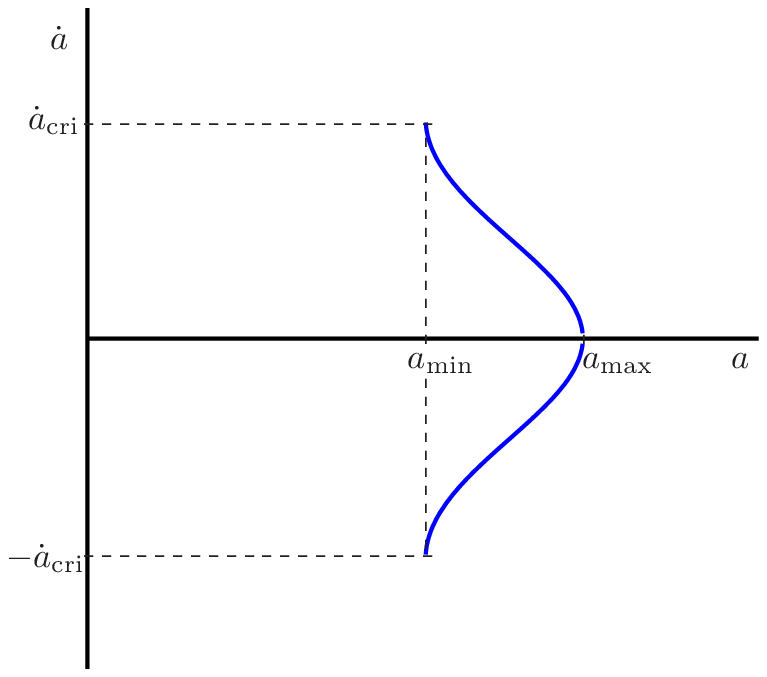}
\centering
\caption{Phase space for $A>0$ and $k=1$ with ``$-$'' sign.\label{fig21}}
\end{figure}

From (\ref{210}) we have 
\begin{equation}
\rho =\frac{a_{0}^{4}}{a^{4}}\rho _{0},
\end{equation}
from where
\begin{equation}
\rho _{\text{min}}=\frac{a_{0}^{4}}{a_{\text{max}}^{4}}\rho _{0}
\end{equation}%
and therefore%
\begin{equation}
\rho _{\text{min}}=\left( \frac{6ka_{0}^{2}}{3\alpha l^{2}k^{2}+\kappa
_{5}\rho _{0}a_{0}^{4}}\right) ^{2}\rho _{0}
\end{equation}

\begin{figure}[h]
\includegraphics[width=350\unitlength]{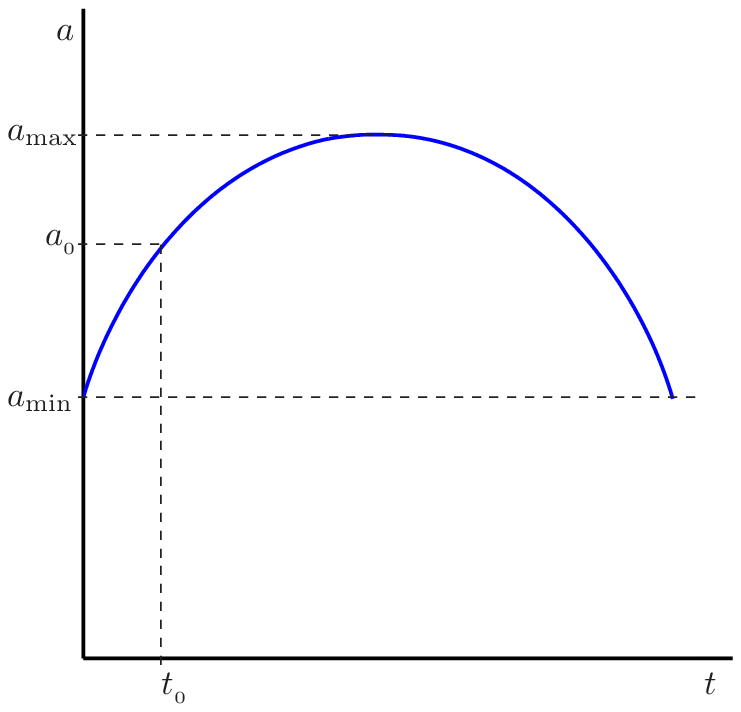}
\centering
\caption{Solution of $\dot a=\sqrt{Aa^2\left(1-\sqrt{1-\frac{a_\text{min}^4}{a^4}} \right)-k}$ with $A>0$, $k=1$ and $\dot a_{_0}<\dot a_\text{cri}$.\label{fig22}}
\end{figure}

.

\subsubsection{Case $\protect\alpha <0$}

In this case 
\begin{equation}
A=\frac{1}{\alpha l^2}<0.
\end{equation}

From  (\ref{eqz41}) we can see that $\dot{a}$ is well defined if 
\begin{equation}
Aa^{2}\left( 1\pm \text{sgn}(A)\sqrt{1-\frac{B}{A}\,\frac{a_{0}^{4}}{a^{4}}}%
\right) -k\geq 0.
\end{equation}%
this constrain exclude the case with plus sign ``$+$'' in front of square root. This condition leads 
\begin{equation}
 a\leq a_{\text{max}}=\sqrt{\frac{%
-ABa_{0}^{4}-k^{2}}{-2Ak}}
\end{equation}%
where $k=1$ and $A=\frac{1}{\alpha l^{2}}<0$. There is a maximum value for $%
a $%
\begin{equation}
a_{\text{max}}=\sqrt{\frac{3\alpha l^{2}k^{2}+\kappa _{5}\rho _{0}a_{0}^{4}}{%
6k}},
\end{equation}%
this maximum leads 
\begin{equation}
\rho _{0}a_{0}^{4}\geq -3\frac{\alpha l^{2}k^{2}}{\kappa _{5}},
\end{equation}%
but (see equation (\ref{eqz30})) 
\begin{equation*}
\rho a^{4}=\rho _{0}a_{0}^{4},
\end{equation*}%
so that,
\begin{equation}
\rho a^{4}\geq -3\frac{\alpha l^{2}k^{2}}{\kappa _{5}}.
\end{equation}%
If there is a maximum $a_{\text{max}}$ then, must exist a minimum for $\rho $%
\begin{equation}
\rho _{\text{min}}=\left( \frac{6ka_{0}^{2}}{3\alpha l^{2}k^{2}+\kappa
_{5}\rho _{0}a_{0}^{4}}\right) ^{2}\rho _{0}.
\end{equation}

\begin{figure}[h]
\includegraphics[width=350\unitlength]{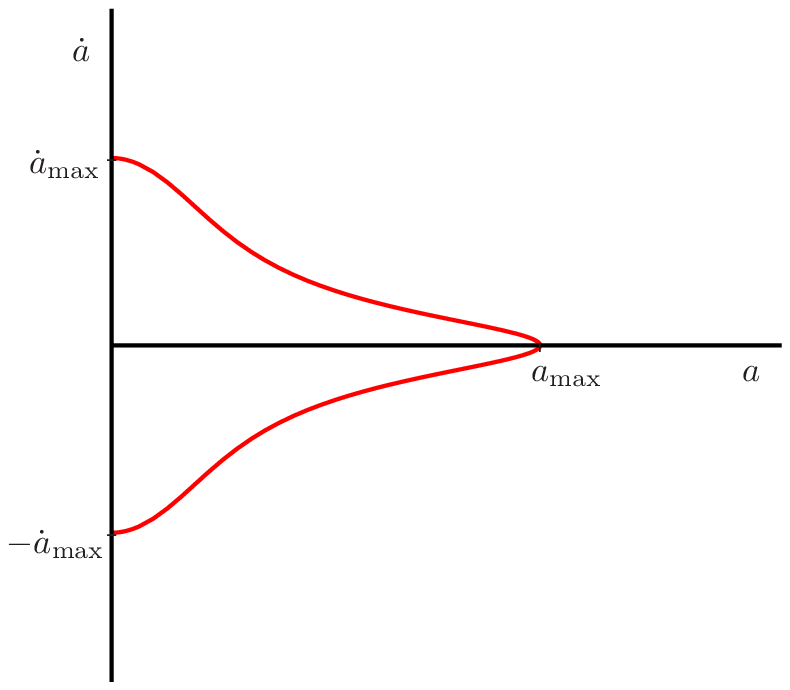}
\centering
\caption{Phase space for $A<0$ and $k=1$ with ``$-$'' sign.\label{fig23}}
\end{figure}

There is no a limit to $a\longrightarrow \infty $ \ and therefore it is
impossible find an approximate solution for

\begin{equation}
\dot{a}=\pm \sqrt{Aa^{2}\left( 1-\sqrt{1-\frac{B}{A}\,\frac{a_{0}^{4}}{a^{4}}%
}\right) -k}.
\end{equation}

\begin{figure}[h]
\includegraphics[width=350\unitlength]{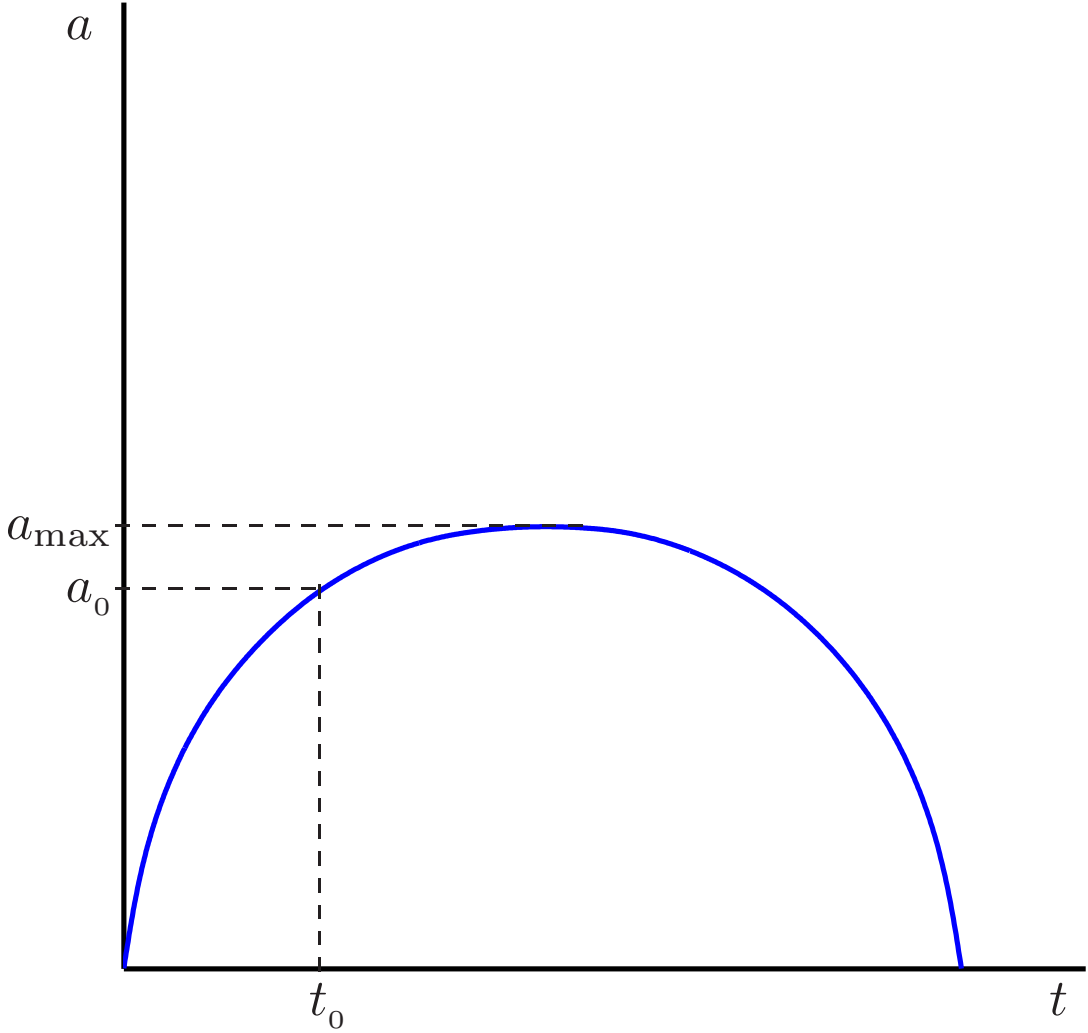}
\centering
\caption{Solution of $\dot a=\sqrt{Aa^2\left(1-\sqrt{1-\frac{Ba_0^4}{Aa^4}} \right)-k}$ with $A<0$, $k=1$ and $\dot a_{_0}<\dot a_\text{max}$.\label{fig24}}
\end{figure}

\subsection{Solutions for era of matter}

We have found a family of solutions for era of matter.\ 

If we consider an open space ($k=-1$), the solutions found include  \begin{inparaenum}[(i)]\item an accelerated expansion ($\alpha>0$) with a minimum scale factor at initial time that, when the time goes to infinity, the scale factor behaves as a hyperbolic sine function (Fig. \ref{fig07}) \item  a decelerated expansion ($\alpha<0$), with a Big Crunch in a finite time $t_\text{max}$ (Fig. \ref{fig09}) \item and a couple of solutions without accelerated expansion, whose scale factor tends to a constant value: $\alpha>0$ (Fig. \ref{fig08}) and  $\alpha<0$ (Fig. \ref{fig11})
\end{inparaenum}
. See Table \ref{table04} and Table \ref{table05}.

\begin{table}[h]
\centering
\caption{Expanding universe solutions for scale factor of  an open space $k=-1$ (hyperbolic) with $\alpha >0$, where \mbox{$a_\text{min}=\sqrt[4]{\frac{\kappa_{5}\alpha l^{2}\rho _{0}}{3}}\,a_{0}$},  \mbox{$\omega=\sqrt{\frac{2}{\alpha l^2}}$}, $\phi=\textrm{arsinh}\left(\frac{\omega}{\sqrt{-k}}a_0\right)$, $\rho _{\max }=\frac{3}{\kappa _{5}\alpha l^{2}}$ and $\dot a_\text{cri}=\sqrt{\sqrt{\frac{\kappa _{5}\rho _{0}}{3\alpha l^{2}}}a_{\text{0}}^{2}-k}$.\label{table04}}
\begin{tabular*}{\columnwidth}{@{\extracolsep{\fill}}l|ll@{}}
\hline
&Accelerated &No accelerated\\
\hline
$a$&$a_\text{min}\leq a$&$a_\text{min}\leq a$\\
$a(t\rightarrow\infty)$&$\sim \sinh\Bigl(\omega(t-t_0)+\phi\Bigr)$&$\sim (t-t_0)$\\
$\rho\sim\frac{1}{a^4}$&$0<\rho\leq\rho_\text{max}$&$0<\rho\leq\rho_\text{max}$\\
$\dot a$&$ \dot a_\text{cri}<\dot a$&$\sqrt{-k}<\dot a<\dot a_\text{cri}$\\
$\dot a(t\rightarrow\infty)$&$\sim \cosh\Bigl(\omega(t-t_0)+\phi\Bigr)$&$\sim \sqrt{-k}$\\
\hline
\end{tabular*}
\end{table}

\begin{table}[h]
\centering
\caption{Expanding universe solutions for scale factor of  an open space $k=-1$ (hyperbolic) with $\alpha <0$, where \mbox{$a_\text{max}=\sqrt{\frac{3\alpha l^{2}k^{2}+\kappa _{5}\rho _{0}a_{0}^{4}}{6k}}$}, \mbox{$\rho _{\text{min}}=\left( \frac{6ka_{0}^{2}}{3\alpha l^{2}k^{2}+\kappa
_{5}\rho _{0}a_{0}^{4}}\right) ^{2}\rho _{0}$} and $\dot a_{\text{max}\pm}=\sqrt{\pm\sqrt{-\frac{\kappa _{5}\rho _{0}}{3\alpha l^{2}}}a_{\text{0}}^{2}-k}$. A decelerated solution describes \textit{Big Crunch} in a finite time $t_\text{max}$.\label{table05}}
\begin{tabular*}{\columnwidth}{@{\extracolsep{\fill}}l|ll@{}}
\hline
&Decelerated &No accelerated\\
\hline
$a$&$0\leq a\leq a_{\text{max}}$&$0\leq a$\\
$a(t\rightarrow\infty)$&$-$&$\sim (t-t_0)$\\
$\rho\sim\frac{1}{a^4}$&$\rho_\text{min}\leq\rho$&$0<\rho$\\
$\dot a$&$ -\dot a_{\text{max}-}\leq\dot a\leq \dot a_{\text{max}-}$&$\sqrt{-k}<\dot a\leq \dot a_{\text{max}+}$\\
$\dot a(t\rightarrow\infty)$&$-$&$\sim \sqrt{-k}$\\
\hline
\end{tabular*}

\end{table}

From models found in Section \ref{sub3.1}  we can see that there are solutions with $\alpha>0$ for accelerated contracting universe and no accelerated contracting  universe (see Figure \ref{fig06}, \mbox{$\dot a<0$}). These solutions were not studied.

Solutions found for a flat universe ($k=0$) in expansion are \begin{inparaenum}[(i)]\item an accelerated expansion whose scale factor behaves as a exponencial function when time grows and starts from a minimum value (Fig. \ref{fig15}) \item and a couple of solutions with decelerated expansion whose scale factor tends to square root function: $\alpha>0$ (Fig. \ref{fig16}) and  $\alpha<0$ (Fig. \ref{fig18})
\end{inparaenum}
. See Table \ref{table06} and Table \ref{table07}.

\begin{table}[h]
\centering
\caption{Expanding universe solutions for scale factor of  a flat space $k=0$ with $\alpha >0$, where \mbox{$a_\text{min}=\sqrt[4]{\frac{\kappa_{5}\alpha l^{2}\rho _{0}}{3}}\,a_{0}$},  $\omega=\sqrt{\frac{2}{\alpha l^2}}$, $\rho _{\max }=\frac{3}{\kappa _{5}\alpha l^{2}}$ and $\dot a_\text{cri}=\sqrt[4]{\frac{\kappa _{5}\rho _{0}}{3\alpha l^{2}}}a_{\text{0}}$. \label{table06}}
\begin{tabular*}{\columnwidth}{@{\extracolsep{\fill}}l|ll@{}}
\hline
&Accelerated &Decelerated\\
\hline
$a$&$a_\text{min}\leq a$&$a_\text{min}\leq a$\\
$a(t\rightarrow\infty)$&$\sim \exp\Bigl(\omega(t-t_0)\Bigr)$&$\sim \sqrt{1+\omega (a_\text{min} /a_0)^2\,(t-t_0)}$\\
$\rho\sim\frac{1}{a^4}$&$0<\rho\leq\rho_\text{max}$&$0<\rho\leq\rho_\text{max}$\\
$\dot a$&$ \dot a_\text{cri}<\dot a$&$0<\dot a<\dot a_\text{cri}$\\
$\dot a(t\rightarrow\infty)$&$\sim \exp\Bigl(\omega(t-t_0)\Bigr)$&$\sim\frac{1}{\sqrt{1+\omega (a_\text{min} /a_0)^2\,(t-t_0)}}$\\
\hline
\end{tabular*}
\end{table}

In this case there are also solutions of contraction universe ($\dot a<0$) \begin{inparaenum}[(i)]\item one ends with a minimum value $a_\text{min}$ when $\alpha$ is positive (Fig. \ref{fig14}) \item and other ends with a Big Crunch when $\alpha$ is negative (Fig. \ref{fig17}).
\end{inparaenum}

\begin{table}[h]
\centering
\caption{Expanding universe solutions for scale factor of  a flat space $k=0$ with $\alpha <0$, where\break $a_\text{ref}=\sqrt[4]{-\frac{\kappa_{5}\alpha l^{2}\rho _{0}}{3}}\,a_{0}$,  $\omega=\sqrt{-\frac{2}{\alpha l^2}}$, and $\dot a_\text{max}=\sqrt[4]{-\frac{\kappa _{5}\rho _{0}}{3\alpha l^{2}}}a_{\text{0}}$.\label{table07}}
\begin{tabular*}{\columnwidth}{@{\extracolsep{\fill}}l|l@{}}
\hline
&Decelerate\\
\hline
$a$&$0\leq a$\\
$a(t\rightarrow\infty)$&$\sim \sqrt{1+\omega (a_\text{ref} /a_0)^2\,(t-t_0)}$\\
$\rho\sim\frac{1}{a^4}$&$0\leq\rho$\\
$\dot a$&$0<\dot a\leq \dot a_\text{max}$\\
$\dot a(t\rightarrow\infty)$&$\sim \frac{1}{\sqrt{1+\omega (a_\text{ref} /a_0)^2\,(t-t_0)}}$\\
\hline
\end{tabular*}
\end{table}

Finally, we only found  one solution for a closed universe ($k=1$) in expansion. This solution is found when $\alpha$ is greater than zero. It behaves as a hyperbolic cosine function when time grows and starts from a minimum value (Fig \ref{fig20}). See Table \ref{table08}.

\begin{table}[h]
\centering
\caption{Expanding universe solutions for scale factor of  a closed space $k=1$ with $\alpha >0$, where $a_\text{min}=\sqrt[4]{\frac{\kappa_{5}\alpha l^{2}\rho _{0}}{3}}\,a_{0}$, $a_\text{max}=\sqrt{\frac{3\alpha l^{2}k^{2}+\kappa _{5}\rho _{0}a_{0}^{4}}{6k}}$,  $\omega=\sqrt{\frac{2}{\alpha l^2}}$, $\phi=\textrm{arcosh}\left(\frac{\omega}{\sqrt k}\,a_0\right)$, $\rho _\text{min}=\left( \frac{6ka_{0}^{2}}{3\alpha l^{2}k^{2}+\kappa
_{5}\rho _{0}a_{0}^{4}}\right) ^{2}\rho _{0}$, \mbox{$\rho_\text{max}=\frac{3}{\kappa _{5}\alpha l^{2}}$} and $\dot a_\text{cri}=\sqrt{\sqrt{\frac{\kappa _{5}\rho _{0}}{3\alpha l^{2}}}a_{0}^2-k}$. A decelerated solution describes an expanding universe, which then stops the expansion and then contracts until scale factor reaches a minimum $a_\text{min}>0$, in a finite time $t_\text{max}$.\label{table08}}
\begin{tabular*}{\columnwidth}{@{\extracolsep{\fill}}l|ll@{}}
\hline
&Accelerated &Decelerated\\
\hline
$a$&$a_\text{min}\leq a$&$a_\text{min}\leq a\leq a_\text{max}$\\
$a(t\rightarrow\infty)$&$\sim \cosh\Bigl(\omega(t-t_0)+\phi\Bigr)$&$-$\\
$\rho\sim\frac{1}{a^4}$&$0<\rho\leq\rho_\text{max}$&$\rho_\text{min}\leq\rho\leq\rho_\text{max}$\\
$\dot a$&$ \dot a_\text{cri}<\dot a$&$-\dot a_\text{cri}<\dot a<\dot a_\text{cri}$\\
$\dot a(t\rightarrow\infty)$&$\sim \sinh\Bigl(\omega(t-t_0)+\phi\Bigr)$&$-$\\
\hline
\end{tabular*}
\end{table}

Furthermore, there are two contracting universe solutions, both ends in a finite time  \begin{inparaenum}[(i)]\item one ends with a minimun value $a_\text{min}$, when $\alpha$ is positive (Fig. \ref{fig22}) \item and other ends with a Big Crunch, when $\alpha$ is negative (See Table \ref{table09} and Fig. \ref{fig24}). 
\end{inparaenum}

\begin{table}[h]
\centering
\caption{Expanding universe solutions with Big Crunch for scale factor of  a closed space $k=1$ with $\alpha <0$, where\break $a_\text{max}=\sqrt{\frac{3\alpha l^{2}k^{2}+\kappa _{5}\rho _{0}a_{0}^{4}}{6k}}$, $\rho _{\text{min}}=\left( \frac{6ka_{0}^{2}}{3\alpha l^{2}k^{2}+\kappa
_{5}\rho _{0}a_{0}^{4}}\right) ^{2}\rho _{0}$ and $\dot a_{\text{max}}=\sqrt{\sqrt{-\frac{\kappa _{5}\rho _{0}}{3\alpha l^{2}}}a_{\text{0}}^{2}-k}$. This solution describes a expanding universe, which then stops the expansion and then contracts until a Big Crunch, in a finite time $t_\text{max}$.\label{table09}}
\begin{tabular*}{\columnwidth}{@{\extracolsep{\fill}}l|l@{}}
\hline
&Decelerate\\
\hline
$a$&$a\leq a_\text{max}$\\
$a(t\rightarrow\infty)$&$-$\\
$\rho\sim\frac{1}{a^4}$&$\rho_\text{min}\leq\rho$\\
$\dot a$&$-\dot a_\text{max}<\dot a<\dot a_\text{max}$\\
$\dot a(t\rightarrow\infty)$&$-$\\
\hline
\end{tabular*}
\end{table}

\section{Summary}

We have considered a five-dimensional Eins\-tein-Chern-Si\-mons action $%
S=S_{g}+S_{M}$ which is composed of a gravitational sector and a sector of
matter, where the gravitational sector is given by a Chern-Simons gravity
action instead of the Einstein-Hilbert action and where the matter sector is
given by the so called perfect fluid. We have shown that

\begin{enumerate}[i]
\item The Einstein-Chern-Simons field equations (\ref{2} - \ref{5}) subject to the conditions $T^{a}=0$, $%
k^{ab}=0$ and $\frac{\delta L_{M}}{\delta \omega ^{ab}}=0$ are re-written in
a way similar to the Einstein Maxwell field equations (\ref{9} - \ref{11}). In the case where the equations (\ref{9} - \ref{11}) satisfy the cosmological
principle and the ordinary matter is negligible compared to the dark energy, we find that the equations (\ref{9} - \ref{11}) take the form (\ref{eqz06} - \ref{eqz10}). When ordinary matter
is modeled as dust (Era of Matter), we find that the equations (\ref{9} - \ref{11}) take the form (\ref{eqz23} - \ref{eqz27}).

\item The field equations (\ref{eqz06} - \ref{eqz10}) were completely resolved for the age of Dark
Energy (Sec. \ref{sec03}, accelerated expansion). We find that the field $h^{a}$ has a similar behavior to that of a cosmological constant.

\item The field equations (\ref{eqz23} - \ref{eqz27}) were solved for the era of Matter (Sec. \ref{sec04}). We find several models that are consistent with standard cosmology. The ynamics of the field $h^{a}$ (\ref{eqz27}) was not analyzed because the focus
was placed on the dynamics of the scale factor $a(t)$.
\end{enumerate}

In fact, in Section \ref{sec03} we have found solutions that describes accelerated expansion for the three possible cosmological models of the universe.
Namely, spherical expansion $\left( k=1\right) $, flat expansion $\left(
k=0\right) $ and hyperbolic expansion $\left( k=-1\right) $ when the
constant $\alpha $ is greater than zero. This mean that the
Einstein-Chern-Simons field equations have as a of their solutions an
universe in accelerated expansion. This result allow us to conjeture that
this solutions are compatible with the era of Dark Energy and that the
energy-momentum tensor for the field $h^{a}$ corresponds to a form of
positive cosmological constant.  We have also shown that the EChS field
equations have solutions that allows us to identify the energy-momentum
tensor for the field $h^{a}$ with a negative cosmological constant.

On the other hand, in Section \ref{sec04} we have found a family of solutions for
era of matter. In the case $k=-1$ (open universe), the solutions correspond
to \begin{inparaenum}[(i)]
\item an accelerated expansion ($\alpha >0)$ with a minimum
scale factor at initial time that, when the time goes to infinity, the scale
factor behaves as a hyperbolic sine function
\item a decelerated expansion ($\alpha <0$), with a Big Crunch in a finite time $%
t_{\max }$ 
\item and a couple of solutions without
accelerated expansion, whose scale factor tends to a constant value.
\end{inparaenum} In the case $k=0$ (flat universe), the solutions describing \begin{inparaenum}[(i)]
\item an accelerated expansion whose scale factor behaves as a exponencial
function when time grows and starts from a minimum value 
\item and a couple of solutions with decelerated expansion whose scale factor
tends to square root function. 
\end{inparaenum}
In the case $k=1$ it is found only one solution for a closed universe in expansion, which behaves as a hyperbolic
cosine function when time grows and starts from a minimum value. However there
are two contracting universe solutions, both ends in a finite time. One ends
with a minimun value $a_\text{min}$, \ when $\alpha $ is positive and other
ends with a Big Crunch, when $\alpha $ is negative.

In summary, we have found some solutions for the field equations, which
were obtained from a Lagrangian for a Chern-Simons gravity theory, studied
in Ref. \cite{Izaurieta2009213}. \ One problem with these solutions is that they are
valid only in a five-dimensional space. 

A connection between five-dimensional spacetimes and the four-dimensional
universe could be accomplished by using a procedure, based on the
Kaluza-Klein theory, known as dynamic compactification \cite{andrew}%
, \cite{mohamedi}.  The method consists in considering a
spacetime metric in which the scale factor of the compact space evolves as an inverse power of the radius of the observable universe. In fact the
metric can be written in a convenient way so that it can achieve the
compactness of the fifth dimension. Following refs. \cite{andrew}, \cite%
{mohamedi} we could consider the 5-dimensional metric
\begin{equation}
ds^{2}=-dt^{2}+a^{2}(t)\left[ \left( dx^{1}\right) ^{2}+\left( dx^{2}\right)
^{2}+\left( dx^{3}\right) ^{2}\right] +b^{2}(t)dx^{2},  \label{153}
\end{equation}
and then consider the case when the scale factor $b(t)$ is given by 
\begin{equation}
b(t)=\frac{1}{a^{n}},\text{ \ }n>0.  \label{217}
\end{equation}
where the parameter $n$ must be positive for dynamical compactification to take place.

Substituting (\ref{217}) into the metric (\ref{153}) we have
\begin{equation}
ds^{2}=-dt^{2}+a^{2}(t)\left[ \left( dx^{1}\right) ^{2}+\left( dx^{2}\right)
^{2}+\left( dx^{3}\right) ^{2}\right] +\frac{dx^{2}}{a^{2n}(t)}.
\end{equation}%
Therefore $b$ gets smaller as the radius of our universe $%
a$ becomes bigger.

It is possible to conjecture that the dinamic compactification
procedure could lead, in a certain limit, to the usual results of
the 4-dimensional general relativity (work in progress).

It should be noted that this compactification procedure, (dynamic compactification) can not be directly implemented on the theory, because this gravity theory is a theory based on a Chern-Simons Lagrangian.

In Ref. \cite{alvarez}, subsequently Ref. \cite{anabal1}, \cite{anabal2}, \cite{anabal3} and most recently Ref.\cite{pais} was pointed out that
Chern-Simons theories are connected with some even-dimensional structures known as gauged Wess-Zumino-Witten ($gWZW$) terms. In Refs. \cite{ssv}, \cite{szv}, \cite{sigr}, was shown that a
five-dimensional Chern-Simons action invariant under the generalized Poincare algebra $\mathfrak{B}_{5}$ induces a gauged Wess-Zumino-Witten term containing the four-dimensional Einstein-Hilbert action.

\begin{acknowledgments}
This work was supported in part by FONDECYT Grants 1130653 and by
Universidad de Concepci\'{o}n through DIUC Grant  212.011.056-1.0.
Two of the authors (F.G., C.Q.) were supported by grants from the Comisi\'{o}n Nacional de Investiga\-ci\'{o}n Cien\-t\'i\-fica y Tecnol\'{o}gica
CONICYT and from the Universidad de Concepci\'{o}n, Chile. M.C. was supported by Grant FONDECYT  1121030 and by Direcci\'on de Investigaci\'on de la Universidad del B\'io-B\'io through Grants DIUBB 1210072/R and GI1221407/VBC. S.delC. was supported by Grant FONDECYT  1110230 and by Pontificia Universidad Cat\'olica de Valpara\'iso through Grants PUCV 123.710
\end{acknowledgments}

\appendix

\section{Obtaining equations (\protect\ref{eqz01}-\protect\ref{eqz05}%
)\label{apendixa}}

From equations (28-32) of Ref. \cite{PhysRevD.84.063506} we know that%
\begin{align}
48\alpha _{3}\left( \frac{\overset{\cdot }{a}^{2}+k}{a^{2}}\right) +24\alpha
_{1}l^{2}\left( \frac{\overset{\cdot }{a}^{2}+k}{a^{2}}\right) ^{2}&=\beta
_{1}T_{00},  \label{43}\\
-24\alpha _{3}\left[ \frac{\overset{\cdot \cdot }{a}}{a}+\left( \frac{%
\overset{\cdot }{a}^{2}+k}{a^{2}}\right) \right]\quad\qquad\qquad&\notag\\
-24\alpha _{1}l^{2}\frac{\overset{\cdot \cdot }{a}}{a}\left( \frac{\overset{\cdot }{a}^{2}+k}{a^{2}}%
\right) &=\beta _{1}T_{11} , \label{44}\\
24\alpha_{3}l^{2}\left( \frac{\overset{\cdot}{a}^{2}+k}{a^{2}}\right)
^{2}&=\beta_{2}T_{00}^{(h)},  \label{45}\\
-24\alpha_{3}l^{2}\frac{\overset{\cdot\cdot}{a}}{a}\left( \frac {\overset{%
\cdot}{a}^{2}+k}{a^{2}}\right) &=\beta_{2}T_{11}^{(h)},  \label{46}\\
24\alpha_{3}l^{2}\left( \frac{\overset{\cdot}{a}^{2}+k}{a^{2}}\right) \left[
\left( g-f\right) \frac{\overset{\cdot}{a}}{a}+\overset{\cdot}{g}\right] &=0
\label{47}
\end{align}
where%
\begin{align}
h^{0} & =f(t)\,e^{0}  \label{48} \\
h^{p} & =g(t)\,e^{p}\text{, \ \ \ \ }p=1,...,4.  \label{49}
\end{align}

In this article we have considered $\beta_{1}=\beta_{2}=\kappa.$ Making this
replacement in (\ref{43}-\ref{49}) and dividing it by $8\alpha_{3}$ we
have
\begin{align}
6\left( \frac{\overset{\cdot}{a}^{2}+k}{a^{2}}\right) +\left( \frac {%
\alpha_{1}}{\alpha_{3}}\right) \left[ 3l^{2}\left( \frac{\overset{\cdot }{a}%
^{2}+k}{a^{2}}\right) ^{2}\right] &=\left( \frac{\kappa}{8\alpha_{3}}\right)
T_{00} , \label{51}\\
-8\left[ \frac{\overset{\cdot\cdot}{a}}{a}+\left( \frac{\overset{\cdot}{a}%
^{2}+k}{a^{2}}\right) \right]\qquad\qquad\qquad\qquad&\notag\\
-\left( \frac{\alpha_{1}}{\alpha_{3}}\right) %
\left[ 3l^{2}\frac{\overset{\cdot\cdot}{a}}{a}\left( \frac{\overset{\cdot}{a}%
^{2}+k}{a^{2}}\right) \right] &=\left( \frac{\kappa }{8\alpha_{3}}\right)
T_{11} , \label{52}\\
3l^{2}\left( \frac{\overset{\cdot}{a}^{2}+k}{a^{2}}\right) ^{2}=\left( \frac{%
\kappa}{8\alpha_{3}}\right) T_{00}^{(h)}& , \label{53}\\
-3l^{2}\frac{\overset{\cdot\cdot}{a}}{a}\left( \frac{\overset{\cdot}{a}^{2}+k%
}{a^{2}}\right) =\left( \frac{\kappa}{8\alpha_{3}}\right) T_{11}^{(h)}&
,\label{54}\\
3l^{2}\left( \frac{\overset{\cdot}{a}^{2}+k}{a^{2}}\right) \left[ \left(
g-f\right) \frac{\overset{\cdot}{a}}{a}+\overset{\cdot}{g}\right] &=0.
\label{55}
\end{align}

Consider now the definition of the constants of Section \ref{sec02}
\begin{equation}
\kappa_{5}=\frac{\kappa}{8\alpha_{3}},\qquad\alpha=-\frac{\alpha_{1}%
}{\alpha_{3}}.  \label{56}
\end{equation}
With these constants, equations (\ref{51} - \ref{55}) take the form 
\begin{align}
6\left( \frac{\overset{\cdot}{a}^{2}+k}{a^{2}}\right) -\alpha\left[
3l^{2}\left( \frac{\overset{\cdot}{a}^{2}+k}{a^{2}}\right) ^{2}\right]
=\kappa_{5}T_{00},&  \label{57}\\
-8\left[ \frac{\overset{\cdot\cdot}{a}}{a}+\left( \frac{\overset{\cdot}{a}%
^{2}+k}{a^{2}}\right) \right] +\alpha\left[ 3l^{2}\frac{\overset{\cdot\cdot}{%
a}}{a}\left( \frac{\overset{\cdot}{a}^{2}+k}{a^{2}}\right) \right]
&=\kappa_{5}T_{11} , \label{58}\\
3l^{2}\left( \frac{\overset{\cdot}{a}^{2}+k}{a^{2}}\right) ^{2}=\kappa
_{5}T_{00}^{(h)}, & \label{70}\\
-3l^{2}\frac{\overset{\cdot\cdot}{a}}{a}\left( \frac{\overset{\cdot}{a}^{2}+k%
}{a^{2}}\right) =\kappa_{5}T_{11}^{(h)}, & \label{71}\\
3l^{2}\left( \frac{\overset{\cdot}{a}^{2}+k}{a^{2}}\right) \left[ \left(
g-f\right) \frac{\overset{\cdot}{a}}{a}+\overset{\cdot}{g}\right] =0.&
\label{72}
\end{align}

Replacing now (\ref{70}) in square brackets (\ref{57}), and (\ref{71}) in
square brackets (\ref{58}), and passing those terms on the right side of the
equations, we find

\begin{align}
6\left( \frac{\overset{\cdot}{a}^{2}+k}{a^{2}}\right)
&=\kappa_{5}T_{00}+\kappa_{5}\alpha T_{00}^{(h)},  \label{74}\\
-8\left[ \frac{\overset{\cdot\cdot}{a}}{a}+\left( \frac{\overset{\cdot}{a}%
^{2}+k}{a^{2}}\right) \right] &=\kappa_{5}T_{11}+\alpha\kappa_{5}T_{11}^{(h)}
,\label{75}\\
3l^{2}\left( \frac{\overset{\cdot}{a}^{2}+k}{a^{2}}\right) ^{2}&=\kappa
_{5}T_{00}^{(h)} , \label{76}\\
-3l^{2}\frac{\overset{\cdot\cdot}{a}}{a}\left( \frac{\overset{\cdot}{a}^{2}+k%
}{a^{2}}\right) &=\kappa_{5}T_{11}^{(h)},  \label{77}\\
\left( \frac{\overset{\cdot}{a}^{2}+k}{a^{2}}\right) \left[ \left(
g-f\right) \frac{\overset{\cdot}{a}}{a}+\overset{\cdot}{g}\right] &=0.
\label{73}
\end{align}

Accommodating some signs in Eqs. (\ref{75}) and (\ref{77}), grouping some
terms (Eqs. \ref{74} and \ref{75}), we have

\begin{align}
6\left( \frac{\overset{\cdot}{a}^{2}+k}{a^{2}}\right)& =\kappa_{5}\left(
T_{00}+\alpha T_{00}^{(h)}\right) , \label{78}\\
8\left[ \frac{\overset{\cdot\cdot}{a}}{a}+\left( \frac{\overset{\cdot}{a}%
^{2}+k}{a^{2}}\right) \right]& =-\kappa_{5}\left( T_{11}+\alpha
T_{11}^{(h)}\right),\\
3l^{2}\left( \frac{\overset{\cdot}{a}^{2}+k}{a^{2}}\right) ^{2}&=\kappa
_{5}T_{00}^{(h)},\\
3l^{2}\frac{\overset{\cdot\cdot}{a}}{a}\left( \frac{\overset{\cdot}{a}^{2}+k%
}{a^{2}}\right)& =-\kappa_{5}T_{11}^{(h)},\\
\left( \frac{\overset{\cdot}{a}^{2}+k}{a^{2}}\right)& \left[ \left(
g-f\right) \frac{\overset{\cdot}{a}}{a}+\overset{\cdot}{g}\right] =0.
\label{79}
\end{align}

In Section \ref{sec025} was considered an energy-momentum tensor of the form 
\begin{align}
\tilde{T}_{\mu\nu} & =T_{\mu\nu}+\alpha T_{\mu\nu}^{(h)} \\
& =\textrm{diag}(\rho,p,p,p,p)\notag\\
&\qquad\quad+\alpha\,\textrm{diag}\left(\rho^{(h)},p^{(h)},p^{(h)},p^{(h)},p^{(h)}\right)\\
& =\textrm{diag}\Bigl(\rho+\alpha\rho^{(h)},p+\alpha p^{(h)}\notag\\
&\qquad\qquad,p+\alpha p^{(h)},p+\alpha
p^{(h)},p+\alpha p^{(h)}\Bigr) \\
& =\textrm{diag}\left(\tilde{\rho},\tilde{p},\tilde{p},\tilde{p},\tilde{p}\right)
\end{align}
where%
\begin{align}
T_{\mu\nu} & =\textrm{diag}(\rho,p,p,p,p) \\
T_{\mu\nu}^{(h)} & =\textrm{diag}\left(\rho^{(h)},p^{(h)},p^{(h)},p^{(h)},p^{(h)}\right).
\end{align}
Writing the functions $f$ and $g$ as%
\begin{equation}
f=h(0),\qquad g=h,
\end{equation}
we find that the equations (\ref{78} - \ref{79}) take the form%
\begin{equation}
6\left( \frac{\overset{\cdot}{a}^{2}+k}{a^{2}}\right) =\kappa_{5}\tilde {\rho%
},
\end{equation}%
\begin{equation}
3\left[ \frac{\overset{\cdot\cdot}{a}}{a}+\left( \frac{\overset{\cdot}{a}%
^{2}+k}{a^{2}}\right) \right] =-\kappa_{5}\tilde{p},
\end{equation}
\begin{equation}
\frac{3l^{2}}{\kappa _{5}}\left( \frac{\overset{\cdot }{a}^{2}+k}{a^{2}}%
\right) ^{2}=\rho ^{(h)},
\end{equation}%
\begin{equation}
\frac{3l^{2}}{\kappa _{5}}\frac{\overset{\cdot \cdot }{a}}{a}\left( \frac{%
\overset{\cdot }{a}^{2}+k}{a^{2}}\right) =-p^{(h)},
\end{equation}%
\begin{equation}
\left( \frac{\overset{\cdot }{a}^{2}+k}{a^{2}}\right) \left[ \left(
h-h(0)\right) \frac{\overset{\cdot }{a}}{a}+\overset{\cdot }{\varphi }\right]
=0.
\end{equation}%
which correspond to the equations  (\ref{eqz01} - \ref{eqz05}).


\begin{thebibliography}{9}

\bibitem{Izaurieta2009213} F. Izaurieta, P. Minning, \ A. P\'{e}rez, E. Rodr\'{\i}guez, P. Salgado, Phys.
Lett. B 678 (2009) 213.

\bibitem{izaurieta:123512} F. Izaurieta, E. Rodr\'{\i}guez, P. Salgado, Jour. Math. Phys. 47 (2006)
123512.

\bibitem{izaurieta:073511} F. Izaurieta, A. Perez, E. Rodr\'{\i}guez, P. Salgado, Jour. Math. Phys. 50
(2009) 073511.

\bibitem{Zanelli:2005sa} J. Zanelli, "Lecture notes on Chern-Simons (super)gravities. Second
edition (February 2008), 2005.

\bibitem{irs1} F. Izaurieta, E. Rodriguez, P. Salgado, Lett. Math. Phys. 80
(2007) 127.

\bibitem{Chamseddine1989291} A. H. Chamseddine, Phys. Lett. B 233 (1989) 291.

\bibitem{Chamseddine1990213} A. H. Chamseddine, Nucl. Phys. B 346 (1990) 213.


\bibitem{PhysRevD.84.063506} F. Gomez, P. Minning, P. Salgado, Phys. Rev. D 84, 063506 (2011).

\bibitem{PhysRevD.85.124026} C.A.C. Quinzacara and P. Salgado, Phys. Rev. D 85 (2012) 124026.

\bibitem{weinberg1972gravitation} S. Weinberg, \textit{Gravitation and cosmology: Principles and applications of the general theory of relativity}, John Wiley \& Sons, New York, (1972).


\bibitem{1475-7516-2012-12-005} S. del Campo, JCAP, 1212 (2012) 005.


\bibitem{andrew} K. Andrew, B. Bolen, and C. Middleton, Gen. Rel.
Grav. 39 (2007) 2061.

\bibitem{mohamedi} N. Mohammedi, Phys. Rev. D 65 (2002) 104018.
\bibitem{alvarez} L. Alvarez-Gaum\'{e} and P.H. Ginsparg, Annals Phys. 161
(1985) 423.

\bibitem{anabal1} A. Anabalon, S. Willison and J. Zanelli, Phys. Rev. D 75 (2007) 024009.

\bibitem{anabal2} A. Anabalon, S. Willison and J. Zanelli, Phys. Rev. D 77 (2008) 044019.

\bibitem{anabal3} A. Anabalon, JHEP (2008) 069

\bibitem{pais} P. Mora and P. Pais, Phys. Rev. D 84 (2011) 044058

\bibitem{ssv} P. Salgado, P. Salgado-Rebolledo, O. Valdivia, Phys. Lett. B 728 (2014) 99

\bibitem{szv} P. Salgado, R.J. Szabo, O. Valdivia, Phys. Rev. D 89 (2014) 084077

\bibitem{sigr} S. Salgado, F. Izaurieta, N. Gonzalez, G. Rubio, Phys. Lett. B 732 (2014) 255

\end{thebibliography}

\end{document}